\documentclass[sigconf,screen]{acmart}

\usepackage{graphicx}
\usepackage{comment}
\usepackage{amsmath}
\usepackage{listings}
\usepackage{booktabs}
\usepackage{adjustbox}
\usepackage[skins]{tcolorbox}
\usepackage{xspace}
\usepackage{enumitem}
\usepackage{fontawesome}
\usepackage{subcaption}
\usepackage{minted}
\usepackage{balance}
\usepackage{transparent}

\copyrightyear{2026}
\acmYear{2026}
\setcopyright{cc}
\acmConference[ICSE '26]{2026 IEEE/ACM 48th International Conference on Software Engineering}{April 12--18, 2026}{Rio de Janeiro, Brazil}
\acmBooktitle{2026 IEEE/ACM 48th International Conference on Software Engineering (ICSE '26), April 12--18, 2026, Rio de Janeiro, Brazil}
\acmDOI{10.1145/3744916.3764531}
\acmISBN{979-8-4007-2025-3/26/04}

\begin{document}

\definecolor{amber}{rgb}{1.0, 0.75, 0.0}
\newcommand{\Star}[0]{{\color{amber}\faStar}\xspace}
\newcommand{\Code}[1]{\begin{small}\texttt{#1}\end{small}}
\newcommand{\TODO}[1]{{\textbf{\color{red}[[TODO: #1]]}}}
\newcommand{\SystemName}[0]{\textsf{StarScout}\xspace}
\newcommand{\RQOne}[0]{\textbf{RQ1} (\textsf{Prevalance}): \emph{How prevalent are fake GitHub stars?}\xspace}
\newcommand{\RQTwo}[0]{\textbf{RQ2} (\textsf{Activity Patterns}): \emph{What are the activity patterns of GitHub repositories and accounts with fake star campaigns?}}
\newcommand{\RQThree}[0]{\textbf{RQ3} (\textsf{Repository Charactersitics}): \emph{What are the GitHub repositories with fake star campaigns?}}
\newcommand{\RQFour}[0]{\textbf{RQ4} (\textsf{Promotional Effect}): \emph{To what extent are fake stars effective in promoting the target GitHub repositories?}}

\tcbset{
  my box2/.style={
    enhanced,
    colframe=#1!80,
    colback=#1!5,
  },
}
\newtcolorbox{summary-rq}{
  my box2=black,
  boxrule=1pt,top=3pt,bottom=3pt,left=4pt,right=4pt
}

\settopmatter{printfolios=true}

\title[Six Million (Suspected) Fake Stars on GitHub: A Growing Spiral of Popularity Contests, Spam, and Malware]{Six Million (Suspected) Fake Stars on GitHub: \\A Growing Spiral of Popularity Contests, Spam, and Malware}

\author{Hao He}
\orcid{0000-0001-8311-6559}
\affiliation{%
  \institution{Carnegie Mellon University}
  \city{Pittsburgh}
  \state{PA}
  \country{USA}
}
\email{haohe@andrew.cmu.edu}

\author{Haoqin Yang}
\orcid{0009-0006-8557-8013}
\affiliation{%
  \institution{Carnegie Mellon University}
  \city{Pittsburgh}
  \state{PA}
  \country{USA}
}
\email{haoqiny@andrew.cmu.edu}

\author{Philipp Burckhardt}
\orcid{0000-0002-8408-1391}
\affiliation{%
  \institution{Socket Inc}
  \city{Pittsburgh}
  \state{PA}
  \country{USA}
}
\email{philipp@socket.dev}

\author{Alexandros Kapravelos}
\orcid{0000-0002-8839-8521}
\affiliation{
  \institution{North Carolina State University}
  \city{Raleigh}
  \state{NC}
  \country{USA}
}
\email{akaprav@ncsu.edu}

\author{Bogdan Vasilescu}
\orcid{0000-0003-4418-5783}
\affiliation{%
  \institution{Carnegie Mellon University}
  \city{Pittsburgh}
  \state{PA}
  \country{USA}
}
\email{vasilescu@cmu.edu}

\author{Christian Kästner}
\orcid{0000-0002-4450-4572}
\affiliation{%
  \institution{Carnegie Mellon University}
  \city{Pittsburgh}
  \state{PA}
  \country{USA}
}
\email{kaestner@cs.cmu.edu}

\begin{abstract}
  GitHub, the de facto platform for open-source software development, provides a set of social-media-like features to signal high-quality repositories.
  Among them, the star count is the most widely used popularity signal, but it is also at risk of being artificially inflated (i.e., \emph{faked}), decreasing its value as a decision-making signal and posing a security risk to all GitHub users.
  In this paper, we present a systematic, global, and longitudinal measurement study of fake stars in GitHub.
  To this end, we build \SystemName, a scalable tool able to detect anomalous starring behaviors across all GitHub metadata between 2019 and 2024.
  Analyzing the data collected using \SystemName,
  we find that: (1) fake-star-related activities have rapidly surged in 2024; 
  (2) the accounts and repositories in fake star campaigns have highly trivial activity patterns;
  (3) the majority of fake stars are used to promote short-lived phishing malware repositories; the remaining ones are mostly used to promote AI/LLM, blockchain, tool/application, and tutorial/demo repositories; 
  (4) while repositories may have acquired fake stars for growth hacking, fake stars only have a promotion effect in the short term (i.e., less than two months) and become a liability in the long term.
  Our study has implications for platform moderators, open-source practitioners, and supply chain security researchers.
\end{abstract}



\maketitle

\section{Introduction}
\label{sec:introduction}

\begin{quote}
\emph{
The more any quantitative social indicator is used for social decision-making, the more subject it will be to corruption pressures and the more apt it will be to distort and corrupt the social processes it is intended to monitor.
}
- Donald T.\ Campbell~\cite{campbell1979assessing}
\end{quote}

\begin{figure}
    \centering
    \begin{subfigure}{0.37\linewidth}
        \centering
        \includegraphics[width=\textwidth, clip=true, trim=85 260 350 180]{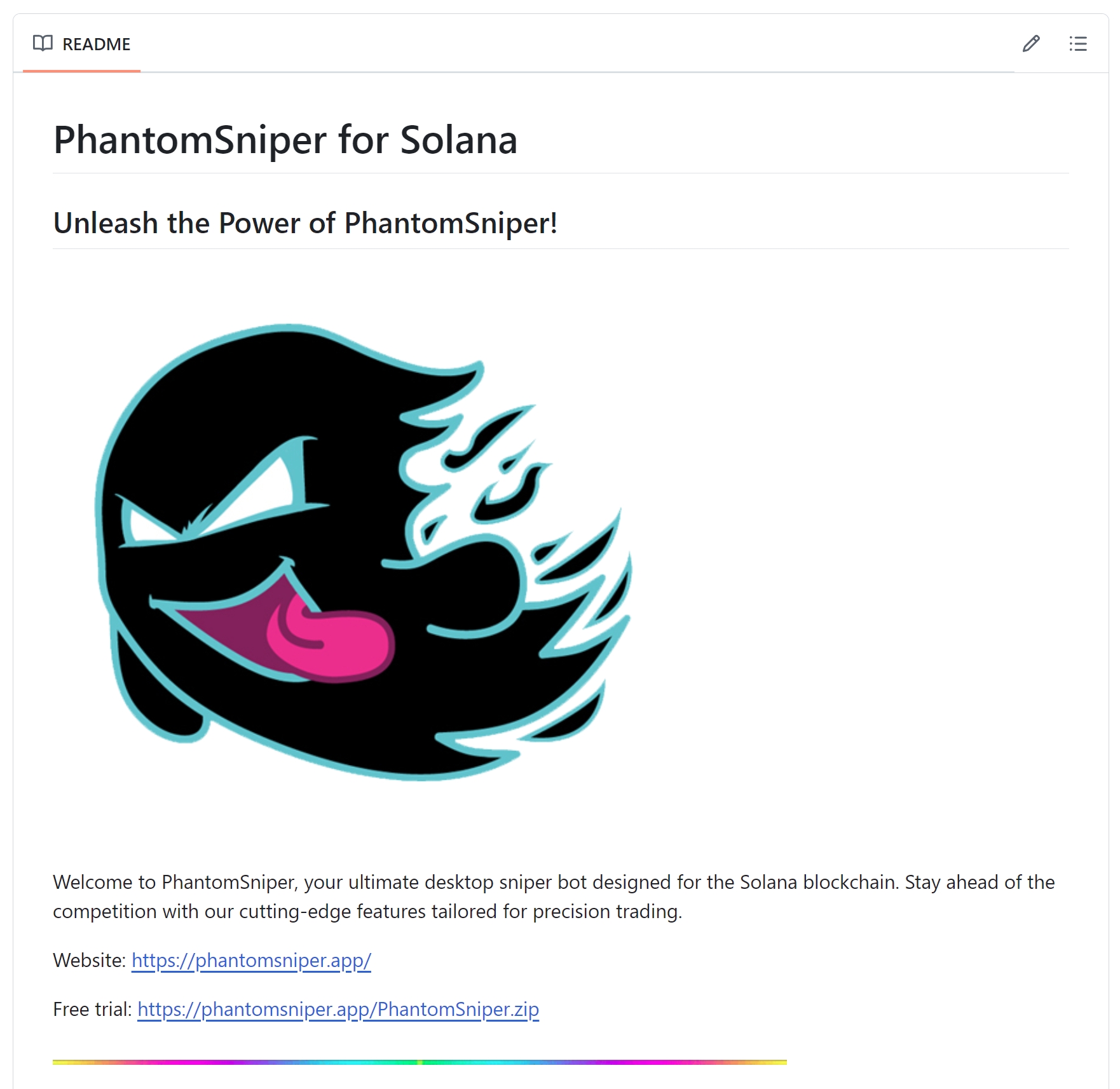}
    \end{subfigure}
    \begin{subfigure}{0.6\linewidth}
        \centering
        \includegraphics[width=\textwidth]{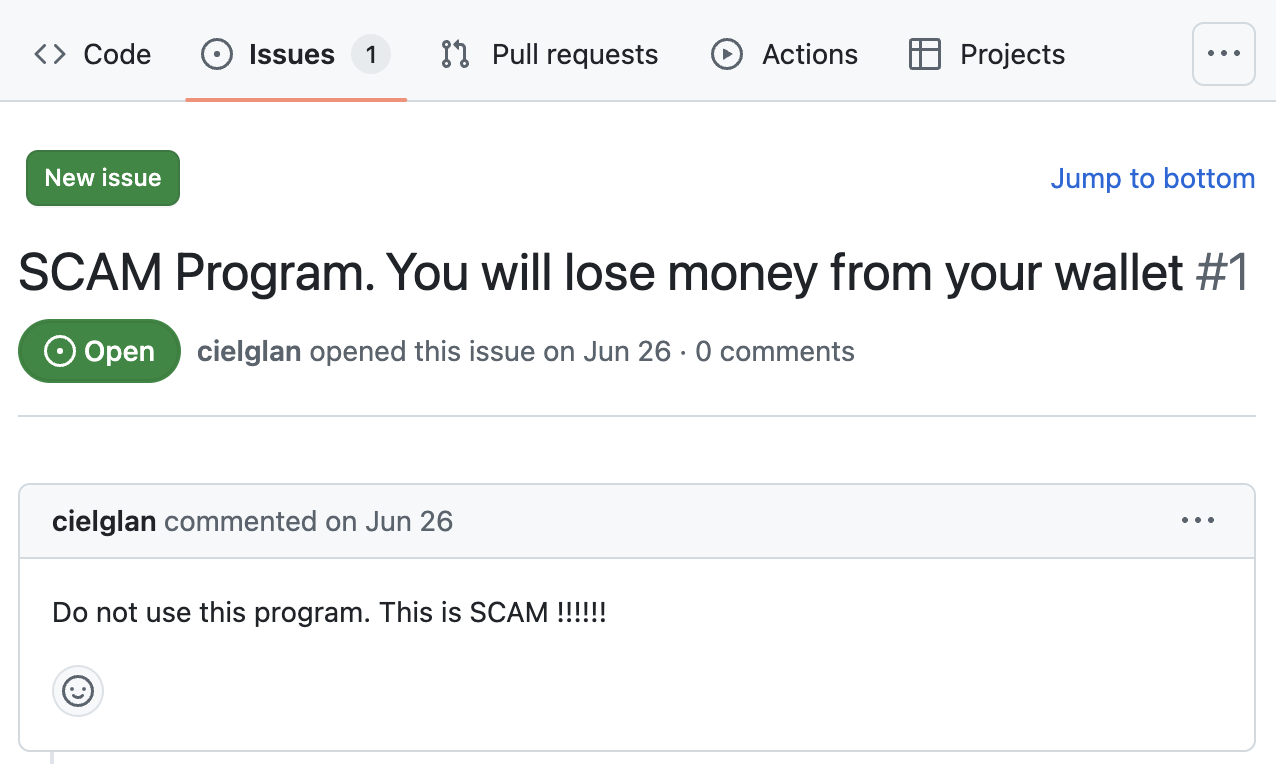}
    \end{subfigure}
\begin{minted}[
frame=lines,
framesep=2mm,
baselinestretch=1,
fontsize=\scriptsize,
]{javascript}
// main.js, hidden at line 12 starting at column 892
const {spawn}=require('child_process');const cmd1='node';
const argv1=['node_modules/@solana-web3-1.43.js'];
spawn(cmd1,argv1,{detached:true,windowsHide:true});
// node_modules/@solana-web3-1.43.js
...;const axios=require('axios'),AdmZip=require('adm-zip'),
fs=require('fs'),{exec}=require('child_process');...;
async function downloadFile(){const _0x3cefa0={'mgLtL':
function(_0x340848,_0x3c32b9){return ...;}}...};
\end{minted}
    \caption{An example malware repository detected by our tool \SystemName, reported to GitHub, and since deleted. It had 111 stars, of which we suspect at least 109 to be fake. The README file (top left) suggests a blockchain application, but if executed, its code (bottom) steals cryptocurrencies using a hidden \Code{spawn()} call to a heavily obfuscated remote file execution script (with the name of a seemingly legitimate JavaScript package). An issue thread (top right), presumably created by a victim, warns of malware hidden inside.}
    \label{fig:malware-example}
\end{figure}

Many studies provided evidence that the number of GitHub stars is widely used when stakeholders evaluate, select, and adopt open-source projects into their software supply chain~\cite{DBLP:journals/jss/BorgesV18, DBLP:journals/pacmhci/QiuLPSV19, DBLP:conf/sigsoft/VargasATBG20, DBLP:journals/ese/MunaiahKCN17,  DBLP:conf/sp/WermkeWKFAF22, koch2024fault}.
However, it is merely a popularity signal (i.e., the attention received from GitHub users), with only moderate correlation to actual usage or importance as shown in prior studies~\cite{koch2024fault, schueller2024modeling, DBLP:journals/corr/abs-2405-07508}. 

More concerning, \emph{GitHub stars can be artificially inflated}~\cite{wired-news}, just as all other popularity signals in social media~\cite{DBLP:journals/compsec/CatalanPVT23}.
For example, a Google search for ``buy GitHub stars'' will reveal a dozen providers (see Appendix, Section~\ref{sec:conclusion}).
From prior reports, we know that GitHub repositories may buy GitHub stars for growth hacking~\cite{oceanbase}, spamming~\cite{dagster}, r\'{e}sum\'{e} fraud~\cite{DBLP:conf/acsac/DuYZDWH0020}, or even for spreading malware~\cite{stargazer-goblin}.
Whatever their purpose, these bought GitHub stars (hereafter ``fake stars'' to match online discussions) corrupt the GitHub star count as a decision-making signal and pose a security threat to all GitHub users (see Figure~\ref{fig:malware-example} for an example).

However, most prior reports of fake GitHub stars come from the gray literature (i.e., non-peer-reviewed blog posts, news reports, and online discussions), and they do not go beyond studying a few specific cases~\cite{dagster, DBLP:conf/acsac/DuYZDWH0020, oceanbase, stargazer-goblin}.
With the rising prevalence of software supply chain attacks~\cite{sonatype-report}, we believe it is important to conduct \emph{a systematic, global, and longitudinal measurement of this emerging threat, especially from the lens of software supply chain security}.
This would contribute to our understanding of fake GitHub stars and the fraudulent/malicious activities around them, furthering the design of countermeasures and informing best practices on the usage of an important software supply chain metric.

Several challenges persist in the global measurement of fake GitHub stars.
First, GitHub stars can be faked in various ways (e.g., bots~\cite{stargazer-goblin}, crowdsourced humans~\cite{oceanbase}, exchange platforms where users exchange stars for a reward~\cite{DBLP:conf/acsac/DuYZDWH0020}), and malicious actors continuously evolve their behaviors to evade platform detection~\cite{DBLP:journals/tissec/IkramOFCFJKS17, DBLP:conf/www/CresciPPST17}.
Their actions blur the boundary between fake and authentic stars, posing a challenge to the identification and measurement of fake ones.
Second, the sheer amount of GitHub data ($\sim$20TB of metadata in the past five years~\cite{gharchive}) and the rate limit of GitHub APIs~\cite{github-rest-api} present challenges to a global analysis.
Finally, GitHub does not preserve deleted repositories and users, posing difficulties in the measurement of fraudulent or malicious activities, as some of them may have already been taken down by the platform.

To overcome these challenges, we take advantage of prior work on mining software repositories~\cite{gharchive} and social media fraud detection~\cite{DBLP:conf/www/BeutelXGPF13, DBLP:conf/www/CresciPPST17} to implement \SystemName, a scalable tool that detects (suspected) \emph{fake stars} over the entire \textsf{GHArchive}~\cite{gharchive} dataset, a Google BigQuery replica of all GitHub events (totaling 63.9 million users, 331 million repositories, 326 million stars, and 6.7 billion other events as of January 2025).
\SystemName (details in Section~\ref{sec:tool-design}) identifies two signatures of anomalous starring behaviors, \emph{the low activity signature} and \emph{the lockstep signature}, and includes a post-processing step to reduce noise and increase overall accuracy.\footnote{
Note that technically speaking, \SystemName only detects \emph{suspected} fake stars and \emph{suspected} fake star campaigns, among which there may still be false positives (see Section~\ref{sec:limitations} for a discussion).
However, to make this paper more readable, we will simply refer to them as \emph{fake stars} and \emph{fake star campaigns} in the remainder of this paper.
}

We apply \SystemName to all GitHub event data from July 2019 to  December 2024, identifying \emph{6.0 million fake stars and 18,617 repositories with fake star campaigns}.
Our evaluation shows that: (a)~\SystemName can detect 81\% of repositories and 76\% of accounts in a confirmed malware campaign involving fake stars~\cite{stargazer-goblin}; (b)~the detected repositories and accounts have anomalously high deletion ratio, up to 90\%, 16x higher compared to random GitHub repositories and accounts.

Using the dataset, we further conduct a measurement study that answers the following research questions:
\begin{itemize}[leftmargin=15pt]
    \item \RQOne
    \item \RQTwo
    \item \RQThree
    \item \RQFour
\end{itemize}

Our findings paint an alarming picture: Fake stars have surged in 2024, either to promote spam or phishing malware repositories, or fuel popularity contests in hyped domains (e.g., blockchain and AI). 
In the latter case, we show that fake stars, while increasing the displayed star counts in the short term, fail to bring true attention in the long term.
From a security perspective, our results call for future research into malicious activities happening on GitHub (e.g., fake stars, fake accounts, spam, phishing).
From a practitioner's perspective, our results further demonstrate the limitation of star counts as a popularity signal, provide counter-evidence to the effectiveness of fake stars for growth hacking, and call for the design of better popularity signals for GitHub repositories.

In summary, this paper makes the following contributions:
\begin{itemize}[leftmargin=15pt]
    \item We design and implement \SystemName, a scalable tool to scan for anomalous starring behavior across all of GitHub.
    This leads to a novel large-scale dataset regarding fake GitHub stars.
    \item Our measurement study discovers novel findings regarding the prevalence, characteristics, and effectiveness of fake stars, leading to practical insights for platform moderators, open-source practitioners, and supply chain security researchers.
\end{itemize}

\section{Background and Related Work}

GitHub is the de facto platform where developers collaborate in open-source projects (organized as \emph{repositories}).
It is designed to be transparent, and repositories can build their reputation through technical and social signals (e.g., stars, forks, badges)~\cite{DBLP:conf/icse/TrockmanZKV18, DBLP:conf/icse/TsayDH14}.
As the design of these social signals resembles those in online social networks, we first introduce spam/fraud research there (Section~\ref{sec:fake-activity}).
Then, we introduce software supply chain attacks (Section~\ref{sec:supply-chain-security}) and summarize prior findings on (fake) GitHub stars (Section~\ref{sec:github-stars}).

\subsection{Fraudulent Activities in Social Networks}
\label{sec:fake-activity}

In a nutshell, most of the fraudulent activities in online social networks are about controlling a group of fake accounts to spread spam or generate fake signals.
There is a rich literature on countering fraudulent activities in online social networks (see, e.g., recent literature reviews~\cite{DBLP:journals/eswa/Latah20, DBLP:journals/eswa/RaoVB21, DBLP:journals/snam/AljabriZSASA23}).
For example, researchers have studied how fake accounts can be used to spread advertisements~\cite{DBLP:conf/imc/ThomasGSP11}, malware~\cite{DBLP:conf/ccs/GrierTPZ10}, misinformation~\cite{shao2018spread}, and political opinions~\cite{faris2017partisanship, pierri2020investigating}.
They can also be used to artificially boost the popularity of posts or users (using, e.g., fake Likes in Facebook~\cite{DBLP:conf/www/BeutelXGPF13, DBLP:journals/tissec/IkramOFCFJKS17} and Instagram~\cite{DBLP:conf/websci/SenAMSKD18}, and fake followers and retweets in Twitter~\cite{DBLP:conf/imc/StringhiniWEKVZZ13, DBLP:conf/ccs/SongLK15}).
For the latter scenario, a recurring observation from prior research is that some providers generate fake popularity signals using automated bots (aka. sybils), but some others use realistic accounts and even crowdsourced real humans (aka. crowdturfing)~\cite{DBLP:conf/imc/StringhiniWEKVZZ13, DBLP:conf/www/WangWZZMZZ12,DBLP:journals/tkdd/YangWWGZD14, DBLP:conf/ccs/SongLK15, DBLP:journals/tissec/IkramOFCFJKS17}. 

With a ground truth dataset and reasonable feature engineering, it is often possible to build high accuracy supervised machine learning models to detect such activities~\citep[e.g.,][]{DBLP:conf/imc/StringhiniWEKVZZ13, DBLP:conf/ccs/SongLK15, DBLP:conf/ccs/XiaoFH15}, but these models can be evaded by malicious actors if they change their behaviors over time~\cite{DBLP:conf/uss/WangWZZ14, DBLP:conf/www/CresciPPST17}.
Another line of research uses unsupervised techniques to detect anomalous users and activity clusters that deviate significantly from normal ones~\cite{DBLP:conf/www/BeutelXGPF13, DBLP:conf/ccs/CaoYYP14, DBLP:conf/uss/ViswanathBCGGKM14, DBLP:conf/kdd/HooiSBSSF16}.
However, \citet{DBLP:journals/tissec/IkramOFCFJKS17} show that the Facebook Like farms are still able to maneuver through these techniques by mimicking realistic user activities and violating the techniques' built-in assumptions (e.g., assuming that fake Likes happen in bursts~\cite{DBLP:conf/www/BeutelXGPF13}).
In general, the battle against fraudulent activities in social media is a never-ending arms race~\cite{DBLP:conf/www/CresciPPST17, yang2019arming}.
Our study contributes to this literature by providing a systematic report of an emerging fraudulent activity (i.e., fake stars) in a novel setting (i.e., GitHub, the de facto social coding platform for open-source projects).

\subsection{Software Supply Chain Attacks}
\label{sec:supply-chain-security}

Modern software development is powered by a collaboratively developed digital infrastructure of open-source software components~\cite{eghbal2016roads}, 
most of which are maintained on GitHub and published in package registries (e.g., npm and PyPI).
In recent years, malicious actors have been trying to exploit both the open-source components and the platforms hosting them, creating an emerging attack vector, often referred to as \emph{software supply chain attacks} (see recent literature reviews~\cite{DBLP:conf/dimva/OhmPS020, DBLP:conf/sp/LadisaPMB23} for more details).
The end goal of these attacks is to sneak malware in a software project,  through which attackers can compromise production systems~\cite{wolff2021navigating}, inject backdoors~\cite{xz-attack}, steal cryptocurrencies~\cite{crypto-attack}, etc.

One common type of attack is to compromise existing open-source components and inject malware into them.
This type of attack can be extremely impactful as open-source components form complex dependency networks in which one component may end up being depended upon by a huge number of downstream applications~\cite{DBLP:conf/uss/ZimmermannSTP19, DBLP:journals/ese/DecanMG19}.
Alternatively, attackers can create new, seemingly useful, but malicious software, and try to get them adopted.
Examples of this kind of attack include typosquatting in package registries~\cite{DBLP:conf/eurosp/VuPMPS20}, publishing malicious VS Code extensions~\cite{vs-code}, and phishing in GitHub repositories~\cite{cao2022fork, stargazer-goblin}.
Fake stars can be a useful weapon for both kinds of attacks: 
An attacker can either establish a fake reputation and seek maintainership of critical open-source projects or buy fake stars to promote their own phishing repositories.
Our study aims to reveal the achieved or potential role of fake stars in (possibly ongoing) software supply chain attacks.

\subsection{GitHub Stars}
\label{sec:github-stars}

Similar to Likes in social networks, GitHub provides a ``Star'' button for users to show appreciation or bookmark a GitHub repository~\cite{DBLP:journals/software/BegelBS13a, DBLP:journals/jss/BorgesV18}.
The widespread use of this feature makes the number of stars a widely recognized popularity signal for GitHub repositories,
e.g., is often used for open-source project selection by both practitioners~\cite{DBLP:journals/pacmhci/QiuLPSV19, DBLP:conf/sigsoft/VargasATBG20, DBLP:conf/sp/WermkeWKFAF22}
and researchers~\cite{DBLP:journals/ese/MunaiahKCN17, koch2024fault}.
However, researchers have also shown that the number of GitHub stars does not correlate highly with the number of downloads and actual installations~\cite{koch2024fault} and does not predict the importance or sustainability of an open-source project~\cite{schueller2024modeling, DBLP:journals/corr/abs-2405-07508}.
Despite studies arguing the limited reliability of GitHub stars, the lack of obvious alternatives makes it still an important signal in these decision-making processes.

\textbf{The GitHub Star Black Market.}
The widespread use of GitHub stars catalyzed the emergence of a GitHub star black market~\cite{wired-news},
which operates similarly to other black markets in social media that sell Twitter followers~\cite{DBLP:conf/imc/StringhiniWEKVZZ13}, Facebook likes~\cite{DBLP:conf/www/BeutelXGPF13}, etc. 
From the gray literature, we know that the GitHub star black market operates in at least three different ways:
(1)~Merchants can sell GitHub stars in batches of (usually) 50 or 100 stars on their own websites (examples in Appendix, Section~\ref{sec:conclusion}), instant messaging apps~\cite{DBLP:conf/acsac/DuYZDWH0020}, or in e-commerce platforms such as Taobao~\cite{taobao-github-star};
(2)~GitHub users may form exchange platforms (e.g., GitStar~\cite{gitstar} or instant messaging groups), where they incentivize mutual starring of their GitHub repositories~\cite{DBLP:conf/acsac/DuYZDWH0020};
(3)~A GitHub repository may directly incentivize the audience of its advertising campaign with gifts if they star the repository (e.g., as it happened in OceanBase~\cite{oceanbase}).
All these ways of operation violate the GitHub Acceptable Use Policy~\cite{github-policy}, which prohibits any inauthentic interactions, rank abuse, and reward-incentivized activities. 
In all three cases discussed above, we consider these purchased, exchanged, or incentivized GitHub stars as \emph{fake} because they are artificially created and do not genuinely represent any authentic appreciations, uses, or bookmarks of a repository from real GitHub users. 

\textbf{Prior Work on Fake GitHub Stars.}
As the first and only academic report on this emerging black market (to the best of our knowledge), Du et al.~\cite{DBLP:conf/acsac/DuYZDWH0020} collected 1,023 black market GitHub accounts through honeypots and built machine learning classifiers to identify 63,872 more suspected GitHub accounts from 2015 to 2019.  
They further conducted a characterization study of these suspected black market fake accounts, but they did not conduct any measurement or characterization of \emph{repositories with fake GitHub stars}. 
On the other hand, the gray literature (i.e., non-peer-reviewed news reports, blog posts, and online discussions) provides evidence on how and why repositories may fake GitHub stars.
For example, repositories may fake GitHub stars for growth hacking~\cite{bohnsack2019hack} (i.e., \emph{``fake it until you make it''}).
Startups (and occasionally, even large companies) buy GitHub stars to promote their open-source products~\cite{oceanbase, dagster} because VCs (and managers) refer to GitHub stars for product viability~\cite{
startup-stars}.
Another reported reason for faking GitHub stars is r\'{e}sum\'{e} fraud, where software developers may want to fake GitHub profiles to increase their competitiveness in hiring~\cite{DBLP:conf/acsac/DuYZDWH0020}.
Finally, malicious actors may fake GitHub stars to promote repositories with malware~\cite{DBLP:conf/www/TaniaMRZF24, stargazer-goblin} (e.g., Figure~\ref{fig:malware-example}).
These reports plot an alarming rising threat to software supply chain security: 
Repositories with fake stars may gain an unfair advantage in the GitHub popularity contest, which can be then exploited in various ways to harm stakeholders in the software supply chain.
To counter this rising threat, it is vital to have a thorough and systematic investigation of fake GitHub stars and the repositories around them.
This forms the main motivation of our study.

\section{\SystemName Design}
\label{sec:tool-design}

\subsection{Overview}

\begin{figure}[t]
    \centering
    \includegraphics[width=\linewidth]{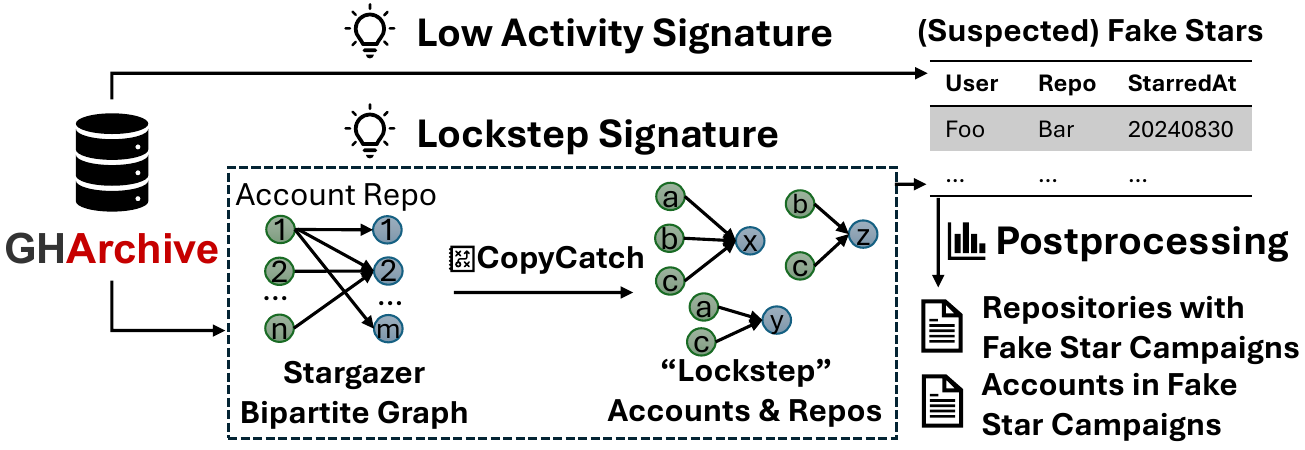}
    \caption{A high-level overview of \SystemName.}
    \label{fig:system-overview}
\end{figure}

At a high level, \SystemName finds two signatures of anomalous starring behaviors from GitHub stargazer history: the \emph{low activity signature} and the \emph{lockstep signature}.
Both signatures are hard to avoid for accounts controlled by GitHub star merchants: Whatever obfuscation methods they use or however realistic they are, these accounts must be either newly-registered throw-away accounts or have been synchronously starring repositories in short time windows to meet their promised delivery times. 
Specifically, the low-activity signature identifies stars coming from accounts that become stale after merely starring one (or a few) repositories; and the lockstep signature identifies stars coming from clusters of $n$ accounts that have been repeatedly acting together to star another cluster of $m$ repositories in short $\Delta t$ time windows. 
However, both signatures may lead to false positives (both to repositories and to accounts), as a legitimate user may happen to behave similarly, or a fake account may star legitimate repositories to hide themselves.
Therefore, \SystemName applies an additional postprocessing step to identify repositories and accounts with fake star campaigns.
We provide an overview of \SystemName in Figure~\ref{fig:system-overview}, and we will describe \SystemName in more detail in the remainder of this section.

\begin{figure}[t]
    \centering
    \includegraphics[width=0.97\linewidth]{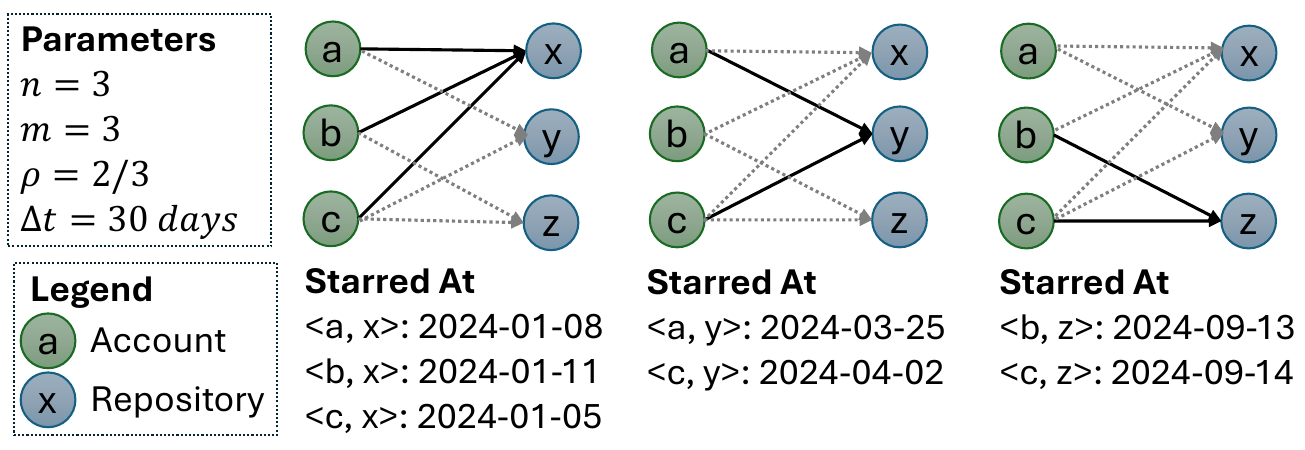}
    \caption{Under the parameter setting shown above, accounts $a, b, c$ and repositories $x, y, z$ are considered to demonstrate a lockstep signature because each repository has stars from at least two of these accounts in a 30-day time window.}
    \label{fig:example}
\end{figure}

\subsection{Heuristics and Postprocessing}
\label{sec:approach}

\textbf{The Low Activity Signature.}
To detect the low activity signature, \SystemName finds GitHub accounts with only one \Code{WatchEvent} (i.e., have only starred one GitHub repository), plus at most one additional event (e.g., \Code{ForkEvent}) in the same repository on the same day.
This heuristic is inspired by a heuristic proposed in the gray literature~\cite{dagster}, but further simplified and tuned to avoid costly interactions with the rate-limited GitHub APIs and easy-to-obfuscate rules. 
While the detected accounts may be throw-away bot accounts controlled by fake star merchants, they may also be real users wrongly identified (e.g., someone legitimately registered an account, starred a repository, and then gave up using GitHub).
To partially address such cases, we also look at fake star counts at the repository level. 
That is, we only consider repositories with at least 50 stars suspected as fake, based on the observation that fake star merchants often sell a minimum of 50 stars (see Appendix, Section~\ref{sec:conclusion}).
Consequently, we ignore the fake stars (and corresponding users) in repositories with fewer than 50 fake stars in total.
We will also use this cutoff of 50 in the lockstep signature detector and the postprocessing step, based on the same intuition.

\textbf{The Lockstep Signature.}
If a fake star merchant controls a group of accounts to repeatedly deliver stars to repositories in their promised delivery time, it would create a unique ``lockstep'' signature.
Prior social network research shows that this signature is unlikely to form naturally and highly correlated to fraudulent activities~\cite{DBLP:conf/www/PanditCWF07, DBLP:conf/www/BeutelXGPF13, DBLP:conf/kdd/HooiSBSSF16}.
Mathematically speaking, let $\mathbf{U}$ be the set of all GitHub accounts and $\mathbf{R}$ be the set of all GitHub repositories; all GitHub stars effectively form a bipartite graph $\langle \mathbf{U}, \mathbf{R}, \mathbf{E} \subseteq \mathbf{U} \times \mathbf{R}, \tau: \mathbf{E}\mapsto \mathbf{T}\rangle$ where each $\langle u, r \rangle \in \mathbf{E} $ represents a star from account $u$ to repository $r$ and $\tau(\langle u, r \rangle)$ maps to the time of starring.
Given four parameters $\langle n, m, \rho, \Delta t\rangle$, we consider a group of accounts $U\subseteq\mathbf{U}$ and repositories $R\subseteq\mathbf{R}$ to demonstrate a \emph{lockstep signature} if:
\begin{align*}
|U| &\geq n, \quad |R| \geq m \\
\forall r \in R, \exists U_r \subseteq U, \exists t_r \in \mathbf{T}: \quad &|U_r| \geq \rho n,\\
&\forall u \in U_r: \langle u, r \rangle \in \mathbf{E},\\ 
&\forall u \in U_r: \tau(\langle u, r \rangle) \in [t_r, t_r + \Delta t]
\end{align*}
In other words,  we consider a group of accounts and repositories to be ``lockstepping'' if each repository has stars from at least $\rho n$ accounts in a time window no longer than $\Delta t$ (see Figure~\ref{fig:example} for an example). 
Note that this definition is both an extension and a relaxation of bicliques (or complete bipartite subgraphs)~\cite{DBLP:journals/dam/Peeters03}, as the latter are generally very rare in real large-scale bipartite graphs.

To find accounts and repositories with the lockstep signature at scale, \SystemName implements \textsf{CopyCatch}~\cite{DBLP:conf/www/BeutelXGPF13}, a state-of-the-art algorithm that has been deployed at Facebook to detect fake Likes with the same lockstep signatures.
At a high level, the algorithm starts from a set of seed repositories.
For each seed repository, it iteratively calibrates a time window and finds the largest possible group of accounts and repositories demonstrating a lockstep signature within that time window.
Finally, groups with more than $n$ accounts and $m$ repositories are retained and considered as engaging in suspicious fake starring activities.
We refer interested readers to the original paper~\cite{DBLP:conf/www/BeutelXGPF13} and our implementation (Section~\ref{sec:data-avail}) for the details of \textsf{CopyCatch}.
While the problem of finding maximum bicliques is NP-complete~\cite{DBLP:journals/dam/Peeters03} and \textsf{CopyCatch} is fundamentally a greedy local search algorithm, it is highly scalable and has been shown to work reasonably well in practice~\cite{DBLP:conf/www/BeutelXGPF13}.

\textbf{Postprocessing.}
Even if the above two signatures identify significantly anomalous patterns in GitHub stargazer data, it is unreasonable to expect that every repository receiving fake stars is actively seeking fake stars. 
In fact, we find that many highly popular repositories get a relatively small amount of fake stars; we infer that fake accounts deliberately star them as an adversarial behavior to evade platform detection.
Therefore, \emph{the postprocessing step aims to identify and keep only repositories with an anomalous spike of fake stars that contributes a non-negligible portion of stars obtained in that repository}.
We infer that these repositories have probably run a fake star campaign and are probably not a victim of fake stargazers, as they gained a noticeable benefit from those stars.
For this purpose, \SystemName aggregates monthly star counts to find repositories where (1) there is at least one single month in which it gets more than 50 fake stars and the percentage of fake stars exceeds 50\%;\footnote{The magic number 50 follows the same rationale as the low activity signature, that fake star merchants often sell a minimum of 50 stars (see Appendix, Section~\ref{sec:conclusion}).} (2) the all-time percentage of fake stars (relative to all stars) exceeds 10\%. 
\SystemName considers these repositories as \emph{repositories with fake star campaigns} and the accounts that starred in these anomalous spike months as \emph{accounts in fake star campaigns}.

\subsection{Implementation}
\label{sec:implementation}

We implement detectors for the low activity signature and each iteration step of \textsf{CopyCatch} (for detecting the lockstep signature) as SQL queries on Google BigQuery, which enables \SystemName to scale the detection to all GitHub in a distributed manner.
We use Python scripts to dynamically generate and send these queries, and collect  fake stars from each query output into a local MongoDB database (using Google Cloud Storage as relay).
We set the lockstep signature parameters as: $n=50$, $m=10$, $\Delta t = 30\text{\emph{~days}}$, and $\rho = 0.5$; in other words, we seek to find clusters of $\ge$50 accounts and $\ge$10 repositories, such that there exists a 30-day interval for each repository in which it gets $\ge$25 stars from these accounts.
As the stargazer bipartite graph is extremely large (326 million edges from July 2019 to December 2024), we split the graph into interleaving six-month chunks (e.g., Jan - Jun 2024, Apr - Sept 2024).
The chunks are interleaved to avoid missing detections across chunk boundaries, and new chunks can be analyzed on a quarterly basis.

We initially deployed \SystemName in July 2024 when it scanned fake stars in the past five years, totaling more than 20 TB of data from \textsf{GHArchive}.
Then, we updated and merged two additional chunks with the latest cutoff of December 2024.
In total, \SystemName identified \textbf{six million (suspected) fake stars} across 26,254 repositories 
before the postprocessing step; 
among these stars, 1.06 million are identified with the low activity signature and 4.93 million with the lockstep signature.
In the postprocessing step, \SystemName identified 18,617 repositories with fake star campaigns and 301k participating accounts (corresponding to 3.81 million fake stars).

\subsection{Evaluation}
\label{sec:evaluation}

It is challenging to evaluate a fake activity detector like \SystemName.
While honeypots are often used in prior research (e.g., \cite{DBLP:conf/acsac/DuYZDWH0020, DBLP:conf/imc/CristofaroFJKS14, DBLP:journals/dss/CresciPPST15}), they raise ethical concerns~~\cite{DBLP:conf/icissp/WeipplSR16, DBLP:conf/uss/KohnoAL23}: If we were to buy fake stars for a honeypot repository, we would be effectively funding malicious actors and creating spam for other developers.
To alleviate this concern, we evaluate the recall of \SystemName using an alternative ground truth malware phishing campaign arguably powered by fake stars.
Specifically, we use the Stargazer Ghost Network dataset~\cite{stargazer-goblin}, which contains 847 repositories with more than 50 GitHub stars coming from 15,672 highly suspicious GitHub accounts. 
We find that \SystemName can detect 688 (81.23\%) of the 847 repositories and 11,903 (75.95\%) of the 15,672 involved GitHub accounts.
\emph{In other words, \SystemName can achieve up to 82.23\% recall for repositories and 75.95\% recall for accounts when it comes to the detection of fake-star-powered malware campaigns.}
We believe this recall performance is satisfactory, especially considering that the algorithm we use for detecting the lockstep signature, \textsf{CopyCatch}, is a randomized greedy algorithm with no completeness guarantees~\cite{DBLP:conf/www/BeutelXGPF13}.

Precision is inherently harder to evaluate because there is no obvious way to examine whether a star is fake or not.
Instead, we estimate \textit{criterion validity}~\cite{cohen1996psychological}, i.e., the extent to which an outcome is correlated to some other theoretically-relevant outcomes (or criterion).
In our case, we compare \emph{the percentage of GitHub repositories and accounts that have been deleted} (as of January 2025) between a comparison sample and the sample detected by \SystemName.
We choose this criterion based on reports of GitHub actively countering fraudulent and malicious activities~\cite{wired-news}---thus, we should be able to observe a higher deletion rate in the \SystemName sample compared to the baseline, especially if fake stars are also used for other malicious activities (like the Stargazer Ghost Network~\cite{stargazer-goblin}).
Therefore, for all repositories and 10,000 sample accounts (downsampled due to GitHub's API rate limit) found by \SystemName, we check (using the GitHub API) whether they were still accessible at the time of detection. 
We also randomly sampled another 10,000 GitHub repositories with $\ge$50 stars and 10,000 GitHub stargazers in these repositories, to get baseline deletion percentages for comparison.

\begin{table}[t]
    \small
    \centering
    \caption{The fraction of repositories/accounts detected by \SystemName and deleted on GitHub at the time of detection is drastically higher compared to the baseline level obtained from random GitHub repositories and users.}
    \label{tab:evaluation}
    \begin{tabular}{lrr}
    \toprule
      & \multicolumn{2}{c}{\% Deleted}\\
    & w/o Postprocessing & w. Postprocessing\\
    \midrule
    \multicolumn{2}{l}{\textbf{Repositories} (Baseline: 5.03\%)} \\
    ~~Low Activity & 14.38\% & 79.36\%\\
    ~~Lockstep & 82.03\% & 90.70\%\\
    ~~\textbf{All} & \textbf{70.05\%} & \textbf{90.42}\%\\
    \multicolumn{2}{l}{\textbf{Accounts} (Baseline: 3.54\%)} \\
    ~~Low Activity & 19.19\% & 72.29\%\\
    ~~Lockstep & 23.03\% & 48.83\% \\
    ~~\textbf{All} & \textbf{18.77\%} & \textbf{57.07\%} \\
    \bottomrule
    \end{tabular}
\end{table}

\emph{Compared with the baseline deletion ratio (5.03\% for repositories and 3.54\% for users), the deletion ratios for the detected repositories and accounts are substantially higher} (Table~\ref{tab:evaluation}): 90.42\% of repositories and 57.07\% of accounts in fake star campaigns (i.e., with postprocessing in Table~\ref{tab:evaluation}) have been deleted. 
The percentage is lower if we consider all repositories and accounts with fake stars (i.e., without postprocessing in Table~\ref{tab:evaluation}), especially for those detected by the low activity signature, providing support for the effectiveness and necessity of the postprocessing step.
Overall, the results align with our intuition that both signatures may be noisy in identifying individual fake stars but, when combined, effective in identifying repositories with fake star campaigns. 
The deletion ratio of accounts is generally lower than that of repositories, which can be interpreted in multiple ways. For example, it may be that \SystemName is less accurate at identifying sophisticated fake accounts (e.g., those detected by the lockstep signature); it is also possible that GitHub prioritizes taking down malware/phishing repositories, as supported by our RQ3 results (Section~\ref{sec:rq3}).

\subsection{Limitations and Ethical Concerns}
\label{sec:limitations}

The most important limitation of \SystemName is that, while it detects statistically anomalous starring patterns in GitHub, the evidence from \SystemName alone is insufficient for disciplinary action (e.g., for taking down fraudulent repositories and accounts).
The latter would usually require evidence from non-public information (e.g., IP addresses).
In general, it is possible that some of the \SystemName identified repositories and accounts may still be false positives.
Another limitation of \SystemName is the use of ad-hoc heuristics and parameters (notably the 50 stars threshold in many parts of detection).
While it is challenging to run thorough sensitivity tests on these parameters due to the sheer amount of data to be processed, we used our best judgment in choosing these parameters.
Also, considering the generally high face validity of our evaluation and measurement study results (Section~\ref{sec:evaluation} and Section~\ref{sec:measurement-study}), we believe the performance of \SystemName suffices for our research goal---characterizing the landscape of an emerging threat.

The possibility of false positives in \SystemName detections also brings ethical concerns.
Even if \SystemName is $\sim$99\% precise in detecting repositories that bought fake stars, there would still be $\sim$180 false positive repositories in our resulting dataset, and interpreting them as having done so would potentially harm their reputation.
To mitigate such risks, we focus on presenting statistical patterns and avoid giving individual repositories as examples in the paper (unless they are obviously malicious).
We also carefully discussed and noted ethical concerns in the dataset documentation, calling for its responsible use and interpretation  (Section~\ref{sec:conclusion}).

\section{Measurement Study}
\label{sec:measurement-study}

Using the dataset of 18,617 repositories and 301k accounts with fake star campaigns (Section~\ref{sec:implementation}), we conduct a measurement study of fraudulent starring activities in GitHub, that answers the following four research questions (formulated in Section~\ref{sec:introduction}):
\begin{itemize}[leftmargin=15pt]
    \item \RQOne
    \item \RQTwo
    \item \RQThree
    \item \RQFour
\end{itemize}
The first three research questions involve different angles of exploratory analysis of this dataset.
The final research question aims to quantify the impact of fake star campaigns: Are they really effective in attracting more (legitimate) attention just as real stars, or only effective in making repositories seem momentarily popular?

\subsection{RQ1: Prevalence}
\label{sec:rq1}

\textbf{Motivation.} 
While \SystemName has detected an enormous number of repositories and accounts with fake star campaigns, it is not yet clear to what extent these repositories and accounts may affect open-source software development and the software supply chain as a whole.
To understand this, we need to provide evidence on: (a)~the prevalence of fake stars and involved repositories \& accounts relative to regular GitHub activities; (b)~the prevalence of fake-star-related repositories in popular distribution channels (e.g., GitHub Trending~\cite{github-trending} and package registries such as npm and PyPI).

\textbf{Prevalence w.r.t. Overall GitHub Activity.}
For each month in our observation period (July 2019 to December 2024), we compute the following: (a)~the percentage of fake GitHub stars among all GitHub stars in that month; (b)~the percentage of GitHub accounts in fake star campaigns, among all active GitHub users (i.e., at least one \textsf{GHArchive} event) in that month; (c)~the percentage of GitHub repositories with fake star campaigns among all repositories that received $\ge 50$ stars in that month.
Through this operationalization, we can estimate the portion of GitHub activities that may have been faked and the impact of these faked activities.

\begin{figure}
    \centering
    \includegraphics[width=\linewidth]{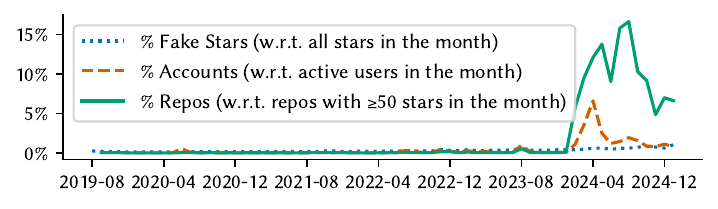}
    \caption{The percentage (\%) of GitHub stars, accounts, and repositories involved in fake star campaigns in each month.}
    \label{fig:prevalence}
\end{figure}

\begin{figure}
    \centering
    \includegraphics[width=\linewidth]{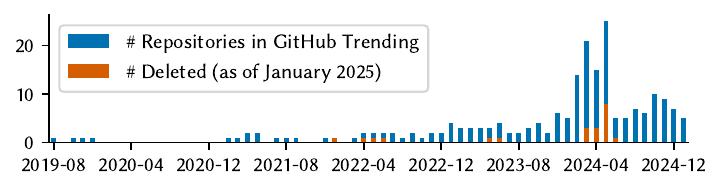}
    \caption{The number (\#) of GitHub repositories with fake star campaigns and appeared in GitHub Trending each month.}
    \label{fig:trending}
\end{figure}

We observe in Figure~\ref{fig:prevalence} that fake stars still only constitute a small fraction ($\le$1\%) of all GitHub stars each month, but fake star campaigns have been increasing since 2022 and surged since 2024.
Their broadest impact was in July 2024, when 16.66\% of popular repositories (3,499) had fake star campaigns. 
March 2024 sees the highest percentage of fake star accounts (6.59\%, 117,024).

\begin{summary-rq}
\textbf{Finding 1:} Fraudulent starring activities started to gain momentum in 2022 and have surged in 2024.
\end{summary-rq}

\textbf{Prevalence in GitHub Trending.}
While the above analysis provides evidence regarding the general prevalence of repositories with fake star campaigns, it is still unclear whether these repositories actually reach developers.
To study this, we pull historical records of GitHub Trending~\cite{github-trending} repositories from a community-maintained archive~\cite{trending-archive}.
Note that this archive only collects trending repositories from five programming languages (Python, Go, C++, JavaScript, and CoffeeScript), so we likely underestimate the actual reach of the studied repositories in GitHub Trending.
In total, we identified 78 (0.42\%) repositories with fake star campaigns that have also appeared in GitHub Trending.
While we also see a similar 2024 surge (Figure~\ref{fig:trending}), the peak number is only 25 (in March 2024).
This lower reach, compared with the results in Figure~\ref{fig:prevalence}, suggests that the GitHub trending algorithm is effective at filtering out most superficially-popular repositories.
Among the matched subset, we anecdotally note repository names suggestive of malicious activities, e.g., \Code{Wallet-stealer} and \Code{blooket-hacks}; we will further analyze these repositories in Section~\ref{sec:rq3}.

\begin{summary-rq}
\textbf{Finding 2:} 
Only a small number (78; 0.42\%) of repositories with fake star campaigns reached the GitHub Trending page.
\end{summary-rq}

\textbf{Prevalence in Package Registries.}
To identify the prevalence of repositories with fake star campaigns in existing package registries, we use the \textsf{ecosyste.ms} API~\cite{ecosystems} to query for explicit references to the GitHub repositories in our dataset.\footnote{This result is also an undercount, as explicit links to GitHub are optional.} 
We matched 738 packages (corresponding to 229 repositories); the most common registries are: PyPI (159 packages), npm (145), Pub (138), Cargo (123), and Go (118).
We further collected their adoption statistics (i.e., \# of dependent packages within the registry and \# of dependent repositories in GitHub).
Overall, results (Table~\ref{tab:downloads}) show scarce
evidence of actual adoption for most of these packages despite their inflated star counts: 70.46\%/77.50\% of these packages do not even have a single dependent open-source package/repository, respectively.

\begin{table}
    \small
    \centering
    \setlength{\tabcolsep}{4pt}
    \caption{Descriptive statistics for the 738 packages linked to 229 repositories with fake star campaigns.}
    \begin{tabular}{lrrrrrrr}
    \toprule
         &  mean & min & 25\% & 50\% & 75\% & max \\
    \midrule
         \# stars & 1686.69 & 54 & 240 & 380 & 1235 & 29.06k\\
         \# dependent packages & 1.69 & 0 & 0 & 0 & 1 & 149 \\
         \# dependent GitHub repos & 5.26 & 0 & 0 & 0 & 0 & 677\\
    \bottomrule
    \end{tabular}
    \label{tab:downloads}
\end{table}

\begin{summary-rq}
\textbf{Finding 3:} 
Only a small number (229; 1.23\%) of repositories with fake star campaigns are also published in package registries (totaling 738 packages). Even fewer are widely adopted.
\end{summary-rq}

\subsection{RQ2: Activity Patterns}
\label{sec:rq2}

\textbf{Motivation.}
While \SystemName detects anomalous starring patterns by construction (Section~\ref{sec:approach}), it does not look into other types of GitHub activities.
Understanding the latter may offer insights into how fake star campaigns operate and what they are used for.
In this RQ, we examine the activity patterns of repositories and accounts with fake star campaigns from two different angles: (a) their overall duration of activity and (b) the distribution of activity types.

\begin{figure}
\centering

\begin{subfigure}{0.49\linewidth}
    \centering
    \includegraphics[width=\linewidth]{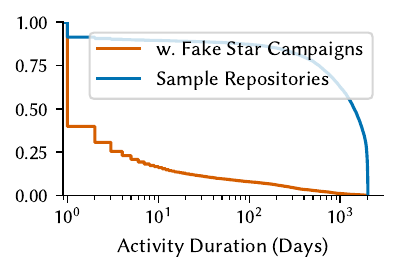}
    \caption{Repositories}
    \label{fig:survival-repos}
\end{subfigure}
\begin{subfigure}{0.49\linewidth}
    \includegraphics[width=\linewidth]{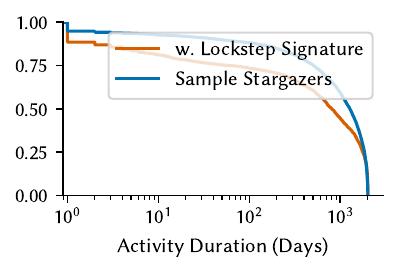}
    \caption{Accounts}
    \label{fig:survival-actors}
\end{subfigure}
\vspace{-1mm}
\caption{The Complementary Cumulative Distribution Function (CCDF) of activity durations for (a) repositories with fake star campaigns and sample repositories, (b) accounts with the lockstep signature and sample stargazers.}
\vspace{-1mm}
\label{fig:survival}
\end{figure}

\textbf{Duration of Activity.}
For repositories and accounts with fake star campaigns, we define their activity duration as the number of days from their first \textsf{GHArchive} event to their last. 
Since accounts with the low-activity signature (31.99\% of accounts in fake star campaigns) have one day of activity by construction, we focus on the activity duration of accounts with the lockstep signature.
For these latter accounts, we compare the Complementary Cumulative Distribution Function (CCDF) of their activity durations with the same comparison samples we used in Section~\ref{sec:evaluation} (Figure~\ref{fig:survival}).
We observe that most repositories with fake star campaigns are short-lived (Figure~\ref{fig:survival-repos}): 83.90\% of repositories with fake star campaigns have less than ten days of activity.
However, user accounts with the lockstep signature can be active for long, with the overall distribution being similar to average GitHub stargazers (Figure~\ref{fig:survival-actors}).

\begin{summary-rq}
\textbf{Finding 4:} 
Most repositories with fake star campaigns are short-lived, but accounts can stay active for a long time. 
\end{summary-rq}

\begin{figure}
\centering
\begin{subfigure}{0.49\linewidth}
    \centering
    \includegraphics[width=\linewidth]{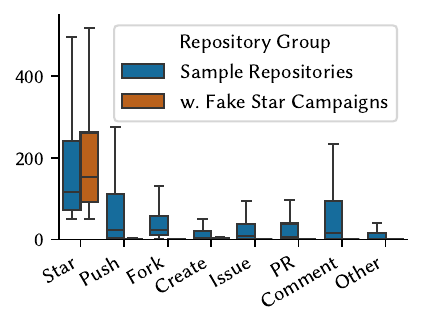}
    \caption{
    Repositories
    }
    \label{fig:repo-events}
\end{subfigure}
\begin{subfigure}{0.49\linewidth}
    \includegraphics[width=\linewidth]{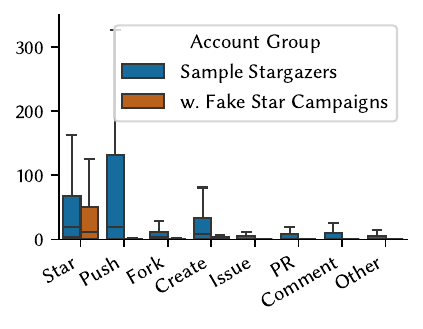}
    \caption{
    Accounts
    }
    \label{fig:actor-events}
\end{subfigure}
\vspace{-1mm}
\caption{Comparing the distribution of GitHub events between: (1) repositories and accounts with fake star campaigns, and (2) random samples of repository and stargazers.}
\vspace{-1mm}
\label{fig:events}
\end{figure}

\textbf{Distribution of Activity Types.}
We aggregate the \textsf{GHArchive} events of repositories and accounts with fake star campaigns into eight types: \emph{Star}, \emph{Push}, \emph{Fork}, \emph{Create}, \emph{Issue}, \emph{PR} (i.e., pull requests), \emph{Comment}, and \emph{Other}.
Then, we compare the overall distributions of activity types with the same samples we used in Section~\ref{sec:evaluation}, aggregating the results in Figure~\ref{fig:events}.
We can observe that repositories and accounts in fake star campaigns tend to be skewed toward starring activities, and they rarely engage in activities that are considerably harder to fake (e.g., issues, pull requests, comments).

However, some repositories and accounts may have different patterns compared to the trends in Figure~\ref{fig:events}. 
To explore this, we run the $k$-Means algorithm~\cite{DBLP:conf/soda/ArthurV07} under different $k$s on normalized activity vectors from repositories and accounts, respectively.
We did not find meaningful clusters for repositories, but we find two additional clusters for accounts: Table~\ref{tab:cluster-center} shows cluster centers for $k = 3$, which has the highest Silhouette Coefficient~\cite{rousseeuw1987silhouettes} among $k \in [2,8]$.
Apart from the majority of accounts that almost exclusively star repositories (Cluster 0, 77.75\%), we find a cluster (Cluster 1, 14.76\%) in which the accounts look generally more realistic---with pushes, repository creations, and other activities.
Our manual analysis of a sample in Cluster 1 indicates that some of them may be creating and pushing artificial repositories to evade platform moderation, but there are also cases where we cannot tell whether they are real or not from their GitHub profiles.
Finally, a small portion of accounts (Cluster 2, 7.49\%) tend to engage in both starring and forking activities; we believe they may come from merchants selling both stars and forks.

\begin{table}
    \centering
    \small
    \caption{The properties of each clustering center identified by $k$-Means on accounts in fake star campaigns.}
    \vspace{-1mm}
    \begin{tabular}{lrrrrrr}
    \toprule
        & \% Star & \% Push & \% Fork & \% Create & \% Other \\
    \midrule
        Cluster 0 (77.75\%) & 95.56\% & 1.00\% &  0.69\% & 2.10\% & 0.65\% \\
        Cluster 1 (14.76\%) & 28.55\% & 40.44\% & 2.55\% & 17.94\% & 10.52\% \\
        Cluster 2 (7.49\%) & 52.34\% & 1.08\% & 43.74\% & 1.76\% & 1.08\%\\
    \bottomrule
    \end{tabular}
    \label{tab:cluster-center}
\end{table}

\begin{summary-rq}
\textbf{Finding 5:} 
Most repositories and accounts with fake star campaigns have trivial GitHub activity patterns.
\end{summary-rq}

\subsection{RQ3: Repository Characteristics}
\label{sec:rq3}

\textbf{Motivation.}
While RQ1 and RQ2 provide statistics on the detected repositories and accounts with fake star campaigns, they do not provide evidence regarding \emph{what exactly these repositories are}.
The anecdotes from the gray literature
suggest that repositories may buy fake stars for popularity contests, growth hacking, or convincing investors~\cite{startup-stars, wired-news}; malicious actors may orchestrate fake star campaigns to spread malware phishing repositories~\cite{stargazer-goblin}.
Can we confirm these reports on \SystemName's detections at a larger scale?
What are the most common domains where fake star campaigns occur?
We answer these subquestions in RQ3.

\begin{table}[t]
    \small
    \centering
    \caption{The most common words in the names of deleted repositories with fake star campaigns.}
    \label{tab:common-words}
    \begin{tabular}{lp{6cm}}
    \toprule
        Category & Words and Occurrences \\
    \midrule
        Pirated software & free (856), crack (721), pro (656), adobe (618), activation (467), cracked (263), studio (239), photoshop (177) \\
        Cryptocurrency & bot (1,071), autoclicker (438), executor (321), crypto (312), wallet (241), trading (175), solana (154)\\ 
        Game cheats & hack (357), cheat (256), roblox (252), h4ck (196)\\
    \bottomrule
    \end{tabular}

\end{table}

\begin{table}[t]
    \small
    \setlength{\tabcolsep}{4.5pt}
    \centering
    \caption{We categorize different groups of repositories with fake star campaigns (trending, packages, and others) into nine categories. Their frequencies are as follows.}
    \begin{tabular}{lrrrr}
    \toprule
       Category & Trending & Packages & Other & All  \\
    \midrule
        \textsf{Spam/Phishing} & 16 (20.5\%) & 22 (9.6\%) & 93 (31.1\%) & 131 (22.6\%)\\
        \textsf{AI/LLM} & 25 (32.1\%) & 36 (15.7\%) & 45 (15.1\%) & 96 (16.6\%)\\
        \textsf{Blockchain} & 10 (12.8\%) & 44 (19.2\%) & 30 (10.0\%) & 76 (13.1\%) \\
        \textsf{Tool/Application} & 15 (19.2\%) & 22 (9.6\%) & 39 (13.0\%) & 73 (12.6\%)\\
        \textsf{Tutorial/Demo} & 4 (5.1\%) & 6 (2.6\%) & 61 (20.4\%) &  70 (12.1\%)\\
        \textsf{Web Frameworks} & 5 (6.4\%) & 44 (19.2\%) & 9 (3.0\%) & 56 (9.7\%) \\
        \textsf{Basic Utility} & 0 (0.0\%) & 27 (11.8\%) & 3 (1.0\%) & 30 (5.2\%) \\
        \textsf{Database} & 0 (0.0\%) & 10 (4.4\%) & 4 (1.3\%) & 14 (2.4\%)\\
        \textsf{Other} & 3 (3.8\%) & 17 (7.4\%) & 15 (5.0\%) & 34 (5.9\%) \\ 
    \bottomrule
    \end{tabular}

    \label{tab:repo-labels}
\end{table}

\textbf{Analysis of Deleted Repositories.}
Our first analysis focuses on the 90.42\% of repositories that have already been deleted on GitHub. 
Unfortunately, it is very difficult to study these repositories: They become inaccessible immediately after deletion from GitHub and \textsf{GHArchive} does not store any repository content.\footnote{
World of Code~\cite{DBLP:journals/ese/MaDBAVTKZM21} (WoC), another possible source of repository data, contains only a small portion of these deleted repositories---we believe they are likely unrepresentative of the remaining ones, depending on the specific discovery mechanisms used in WoC.
}
Despite this, we can still glean some information from a term frequency analysis of their \emph{repository names} (Table~\ref{tab:common-words}). 
We find that most of the deleted repositories carry names indicating pirated software (e.g., \Code{Adobe-Animate-Crack}), cryptocurrency bots (e.g., \Code{pixel-wallet-bot-free}, \Code{Solana-Sniper-Bot}), or game cheats (e.g., \Code{GTA5-cheat}).
While GitHub may have taken some of them down due to copyright violations, it is probably not the majority case: The gray literature shows that fake star campaigns are used to promote phishing malware repositories~\cite{stargazer-goblin}. 
Combining with the evidence we obtained from non-deleted repositories sharing similar names with deleted ones (e.g., Figure~\ref{fig:malware-example}, more examples in Appendix, Section~\ref{sec:conclusion}),
we infer that most of these repositories could be phishing malware repositories masquerading as pirated software, game cheats, and cryptocurrency bots.

\begin{summary-rq}
\textbf{Finding 6:}
Most (90.42\%) repositories with fake star campaigns have been deleted on GitHub. 
Existing evidence suggests that they are probably phishing malware repositories masquerading as pirated software, game cheats, and cryptocurrency bots.
\end{summary-rq}

\textbf{Analysis of Non-Deleted Repositories.}
For the remaining non-deleted repositories (1,783 in total), we clone them immediately at the time of identification. 
Then, we conduct open coding~\cite{khandkar2009open} on the repository READMEs---referring to other files where necessary---on three groups of repositories:
(a) the 78 repositories that appeared in GitHub Trending (Section~\ref{sec:rq1}), which clearly reaches a wide range of developers;
(b) the 229 repositories that are linked by open-source packages (Section~\ref{sec:rq1}), which may have different characteristics compared to others;
(c) 299 randomly sampled repositories from the remaining 1,476 repositories (sample size determined from 95\% confidence level and 5\% margin of error).
From this analysis, we identify eight major categories of repositories (Table~\ref{tab:repo-labels}).
Among them, the most common---and also most concerning---category, is \textsf{Spam/Phishing} (31.1\% in Other, 22.6\% in All), where the repository is either an obvious clickbait (e.g., claiming game cheats, pirated software, cryptocurrency bots) to trick download of suspicious files or spreading spam content (e.g.,  used for search-engine-optimizations).\footnote{Note that while we try our best to identify spam/phishing repositories, it is still possible that we miss better-obfuscated malicious repositories and mis-categorize them into other categories.
We will discuss more on this limitation in Section~\ref{sec:discussion}.}
The remaining popular categories include: (1) \textsf{AI/LLM}, many of which are academic paper repositories or LLM-related startup products, (2) \textsf{Blockchain} or cryptocurrency related repositories, (3) \textsf{Tool/Applications}, or (4) \textsf{Tutorials/Demos} that seem to serve a reference or demonstration purpose (e.g., Python coding examples or list of awesome LLM tools).
We provide definition of each category, analysis of inter-rater reliability, and examples of \textsf{Spam/Phishing} repositories in Appendix (Section~\ref{sec:conclusion}).

\begin{summary-rq}
\textbf{Finding 7:} 
A significant portion ($\sim$30\%) of repositories with fake star campaigns, that are still accessible on GitHub at the time of writing (i.e., March 2025), are spam or active phishing malware repositories. 
The remaining ones are mostly AI/LLM, blockchain, tool/applications, or tutorial/demo repositories.
\end{summary-rq}

\subsection{RQ4: Promotional Effect}
\label{sec:rq4}

\textbf{Motivation.}
For the first three research questions, we have presented a set of exploratory results around the prevalence of fake stars and the characteristics of repositories and accounts with fake star campaigns.
While our RQ3 results highlight that the majority of repositories with fake star campaigns are likely short-lived phishing/spam repositories, there are still a non-negligible number of repositories in our dataset that remain online longer and seem more legitimate.
Why would they buy fake GitHub stars?
The gray literature often argues that (non-malicious) GitHub repositories buy fake stars for growth hacking~\cite{bohnsack2019hack}
(i.e., ``fake it until you make it''), especially because the number of stars is often considered a key indicator of the success of open-source projects and the occasional start-up companies associated with them~\cite{startup-stars}.
Motivated by these speculations, we empirically analyze whether buying fake stars is effective in attracting substantial additional \emph{real} attention, or only effective in making the repositories momentarily seem a little more popular (by the amount of stars they buy).

\textbf{Hypotheses.}
We formulate the following two hypotheses regarding the effect of (fake and real) GitHub stars:
\begin{itemize}[leftmargin=15pt]
    \item \textbf{H$_1$:} \emph{Accumulating real GitHub stars will help a GitHub repository gain more real GitHub stars in the future.}
    \item \textbf{H$_2$:} \emph{Accumulating fake GitHub stars will help a GitHub repository gain more real GitHub stars in the future, but the effect is not as strong as that of real stars.}
\end{itemize}
We formulate \textbf{H$_1$} based on the well-known ``rich get richer'' phenomenon confirmed repeatedly in online social media and similar online contexts~\cite{DBLP:journals/corr/ZhuL16, hagar2022concentration}.
We formulate \textbf{H$_2$} because there is no straightforward way for a GitHub user to distinguish fake stars from real ones while browsing a repository's web page (indeed, we had to build and run a tool for this), and thus fake stars should have a similar effect as real stars for most users.
However, some users may look for other signals (e.g., issues and pull requests) and decide not to star the repository if other signals indicate poor quality, so we expect the effect to be smaller compared to real stars.

\textbf{Methodology.}
To test our hypotheses, we fit \emph{panel autoregression} models~\cite{hanck2021introduction}, estimating the longitudinal effect of gaining fake stars on attracting real stars in the future.
Panel autoregression models are a family of regression models tailored to work on panel data (e.g., variables are collected per repository in a series of time periods) with temporal dependencies (i.e., variables at time $t$ may be affected by variables at time $t - \Delta t$).
By adding fixed effect or random effect terms to the model (we report on both specifications below, to demonstrate that our results are robust), we will be able to estimate the longitudinal effect of independent variables robustly to unobserved heterogeneity (i.e., factors that may affect the outcome variable but are not measured in the model).
Specifically, 
we collect the following variables for each repository $i$ in each month $t$: 
\begin{itemize}[leftmargin=10pt]
    \item $\textit{fake}_{i,t}$: Count of fake stars repository $i$ gained in month $t$;
    \item $\textit{all\_fake}_{i,t}$:  Total count of fake stars in repository $i$ as of $t$;
    \item $\textit{real}_{i,t}$: Count of non-fake stars repository $i$ gained in month $t$;
    \item $\textit{all\_real}_{i,t}$: Total count of non-fake stars in repository $i$ as of $t$;
    \item $\textit{age}_{i,t}$: Number of months since repository $i$ creation;
    \item $\textit{release}_{i,t}$: Whether repository $i$ has at least one GitHub release before or during month $t$;
    \item $\textit{activity}_{i,t}$: The number of \textsf{GHArchive} events for repository $i$ in month $t$, that come from GitHub users who are not the repository owner or one of the fake stargazers. This is a proxy measure for the amount of authentic development activities in month $t$. 
\end{itemize}
The first four variables are intended to test our hypotheses; the remaining three are control variables that have been shown to correlate with star increases in prior work~\cite{DBLP:conf/icse/FangLHV22}. 
We fit $AR(k)$ (i.e., $k$-th order) models for $k = 1, 2, ..., 6$, defined as follows:
\begin{align*}
    \textit{real}_{i,t} \sim &\sum_{j=1}^k \textit{real}_{i, t-k} + \textit{all\_real}_{i, t - k - 1} + \sum_{j=1}^k \textit{fake}_{i, t-k} \\& + \textit{all\_fake}_{i, t - k - 1} + \textit{age}_{i,t} + \textit{release}_{i,t} + \textit{activity}_{i,t}\\
\end{align*}
The last three control variables are for random effect models only; unobserved heterogeneity in fixed effect models is controlled by the per-time and per-repository 
fixed-effect terms.
All variables with skewed distributions are log-transformed before they are fed into the model.
We fit the models using the R \Code{plm} package~\cite{r-plm}.

\textbf{Limitations.}
Before diving into the modeling results, we note the inherent limitations of our modeling approach.
First, even if it allows for robust estimates of coefficients to represent the effect of independent variables, a regression model like ours may still provide insufficient evidence for causality claims.
While Granger causality tests are often used in the econometric literature to establish stronger evidence of causality~\cite{dumitrescu2012testing}, the majority of time series in our data are too short to fulfill the minimum requirements of Granger causality tests on unbalanced panels ($t \ge 6+2k$, $k$ is the autoregression order).
More importantly, it is possible that real causality hides in exogenous variables~\cite{engle1983exogeneity} (e.g., if the repository maintainer is technically inexperienced and shortsighted, that may both cause the maintainer to buy fake stars and the repository to gain fewer real stars).
Therefore, while our results provide initial evidence on the effect of fake stars on GitHub, future work is necessary to establish stronger evidence of causality and further uncover the mechanisms behind such effects.

\begin{table}
    \small
    \centering
    \caption{Fixed/random effect panel autoregression results.}
    \begin{tabular}{lrr} 
    \toprule
     & \multicolumn{2}{c}{\textit{Dependent Variable: $real_{i,t}$}} \\
     & Random Effect & Fixed Effect \\
    \midrule 
     $\textit{real}_{i, t - 1}$ & 0.496$^{***}$ (0.015) & 0.364$^{***}$  (0.018)\\ 
     $\textit{real}_{i, t - 2}$ & 0.187$^{***}$ (0.014) & 0.148$^{***}$  (0.015)\\ 
     $\textit{all\_real}_{i, t - 3}$ & 0.069$^{***}$ (0.008) & 0.097$^{***}$ (0.021)\\ 
     $\textit{fake}_{i, t - 1}$ & 0.058$^{***}$ (0.011) & 0.074$^{***}$ (0.012) \\ 
     $\textit{fake}_{i, t -  2}$ & 0.024$^{**\textcolor{white}{*}}$ (0.010) & 0.029$^{**\textcolor{white}{*}}$ (0.011) \\ 
     $\textit{all\_fake}_{i, t - 3}$ & $-$0.026$^{***}$ (0.006) & $-$0.045$^{***}$ (0.014) \\ 
     $\textit{age}_{i,t}$ & $-$0.004$^{***}$ (0.001)  &  \\ 
     $\textit{release}_{i,t}$ & 0.084$^{***}$  (0.022) &  \\ 
     $\textit{activity}_{i,t}$ & 0.087$^{***}$ (0.010) &  \\ 
    \midrule
    Observations & 12,738 & 12,738 \\ 
    $R^{2}$ & 0.665 & 0.268 \\ 
    Adjusted $R^{2}$ & 0.664 & 0.202 \\ 
    \bottomrule
    \textit{Note:}  & \multicolumn{2}{r}{$^{*}$p$<$0.1; $^{**}$p$<$0.05; $^{***}$p$<$0.01} \\ 
    \end{tabular} 
    \label{tab:regression}
\end{table}

\textbf{Results.}
In this paper, we report results from the $AR(2)$ fixed effect and random effect model in Table~\ref{tab:regression}.
It is worth noting that we obtain consistent findings across all $AR(k)$ models; the remaining results and additional robustness checks are available in the paper Appendix (Section~\ref{sec:conclusion}).
We find strong support for  \textbf{H$_1$}:
According to the fixed-effects model,  a 1\% increase of real stars in month $t-1$ is correlated with an expected 0.36\% increase of real stars in month $t$ 
(since both variables are log-transformed in the model, the coefficient estimates correspond to percentage changes in the outcome)
holding all other variables constant. 
Analogously, one can expect a 0.36\% increase of real stars from month $t$ to month $t+1$.
The effect decreases to 0.15\% 
in month $t+2$ and to 0.10\% 
for all subsequent months, but the effect is always positive.
In other words, a repository with more real stars tends to also get more real stars in the future, echoing the ``rich get richer'' phenomenon prevalent in social networks~\cite{DBLP:journals/corr/ZhuL16, hagar2022concentration}.

On the other hand, \textbf{H$_2$} is only partially supported: while a 1\% increase of fake stars in month $t$ is associated with an expected 0.07\% 
increase of real stars in month $t + 1$ and 0.03\% 
in month $t+2$, holding all other variables constant.
In other words, fake stars do have a statistically significant, longitudinally decreasing positive effect on attracting real stars in the next two months, but the effect is about 5x smaller than that of real stars. 
However, a 1\% increase of fake stars in month $t$ is correlated with an expected 0.04\% 
\emph{decrease} (note the negative coefficient) of real stars on average for all months since month $t+2$.
This suggests that buying fake stars may only help the repository gain new attention in the short term (i.e., less than two months), but the history of faking stars becomes a liability and may result in less overall attention in the long term, perhaps because the repository lacks in other activity and health indicators.

\begin{summary-rq}
\textbf{Finding 8:} Buying fake stars may help a repository gain real attention in the short term (i.e., less than two months), but the effect is about 5x smaller compared to real stars; in the long term, buying fake stars has a negative effect on star gain.
\end{summary-rq}

\section{Discussion: What to Do about Fake Stars?}
\label{sec:discussion}

Our study points to a growing problem: 
Fake star campaigns have become orders of magnitude more common in 2024 compared to 2023, further distorting the already questionable reliability of star count as a signal of popularity and trustworthiness.
Most alarmingly, fake stars are demonstrably associated with increasing security risks and spam/phishing activities. 
In this final section, we discuss potential countermeasures to this problem from the lens of different stakeholders who may take the responsibility.

\subsection{Open-Source Practitioners}

Our study is far from the first showing that stars on GitHub are not a reliable signal~\cite{koch2024fault, github-star-trust, schueller2024modeling, DBLP:journals/corr/abs-2405-07508}. 
We provide further evidence that stars may not only misrepresent popularity or reputation, but can be intentionally manipulated by malicious actors.
Yet, study after study shows how practitioners use 
star counts as an important (and often as the most important) signal for all kinds of decisions, including high-stakes decisions like open-source component adoption~\cite{DBLP:conf/sigsoft/VargasATBG20, DBLP:conf/sp/WermkeWKFAF22,DBLP:journals/jss/MujahidAS23} and repository reputation evaluation~\cite{DBLP:journals/pacmhci/QiuLPSV19, DBLP:conf/icse/TsayDH14}.

We suggest that \emph{star counts should not be used for high-stakes decisions by themselves}.
Yet, we see little hope for practitioners to change this widespread practice at scale any time soon, since it is already deeply ingrained and there are no immediately available and obvious alternatives. 
Still, we hope that exposing the security risks from such reliance may help motivate some changes until better, hopefully empirically validated, signals from platforms or third parties become established (Section~\ref{sec:implication-platform} and~\ref{sec:implications-sca}).

In addition, while it is natural for open source developers to promote and try to attract attention to their projects~\cite{DBLP:conf/msr/FangKLHV20, DBLP:conf/icse/FangLHV22,bohnsack2019hack}, \emph{we recommend against buying fake stars for growth hacking}. 
Not only is it arguably unethical and in violation of GitHub's terms of service, but also our RQ4 results suggest that it is ineffective.
Our interpretation is that there is no plausible mechanism to convert stars into real, sustained adoption, even if a high star count may increase project visibility in the short term.  
Instead, we recommend that \emph{repository maintainers and startup founders in open source should focus on strategies to actually build open-source communities, referring to the vast amount of literature on this topic}~\citep[e.g.,][]{kim2006community, product-approach}. 

\subsection{Platform Providers}
\label{sec:implication-platform}

Platform providers such as GitHub have a lot of leverage for designing features like stars and for designing security mechanisms against their abuse, and they have privileged information such as IP addresses from users that are not publicly available to researchers and third parties.
We believe that \emph{GitHub would benefit from a better designed popularity signal other than raw star counts}. 
The current system, which gives equal weight to all stars, is prone to manipulation from bots and coordinated inauthentic activities.
The literature on review and reputation systems has many ideas for designing more robust measures that can provide inspiration, such as (1)~highlighting signals of real adoption as a popularity measure (e.g., based on dimensions of network centrality~\cite{DBLP:journals/corr/abs-2405-07508}); (2)~incorporating a reputation system where stars of more active or authentic users or of developers actually adopting the dependency are weighted higher~\citep[e.g.,][]{DBLP:conf/asunam/Movshovitz-AttiasMSF13, dellarocas2010online}), and (3)~shadowbanning stars from suspicious users~\cite{risius2024shadowbanning, mukherjee2013yelp}.
Stack Overflow's reputation system and Yelp's review filtering are examples of deliberate designs to combat low quality and malicious user inputs.
Our \SystemName tool is immediately actionable for a shadowbanning approach.

Another key takeaway from our research findings is that \emph{fake stars are highly related to malicious activities and should be considered as such in platform moderation.}
Our manual investigation of repositories in RQ3 reveals multiple cases of malicious repositories that had been stripped of all their stars but still remained accessible on GitHub (examples in Appendix, Section~\ref{sec:conclusion}).
We speculate that GitHub may 
routinely take down known fake accounts or malware repositories, 
but may not link them together.
We suggest that a platform provider could use internal signals of suspicious activity, including fake stars, to direct malware detection efforts.

Finally, our study provides an example of \emph{how platform transparency enables large-scale independent studies of fraud and malicious activities}.
We could not have conducted this study without GitHub's high level of openness, which unfortunately, is not common in many other online platforms.
\emph{We call for the same level of openness in others to enable ``the scrutiny of a thousand eyes.''}

\subsection{Third-Party Tool Providers}
\label{sec:implications-sca}

While the platform itself has the most direct access to affect change for all its users, third parties can also offer valuable services for the entire community. 
They may also be more frank in calling out suspicious practices than a platform owner may be comfortable.

In our case, \emph{software composition analysis (SCA) tool providers are a good match for auditing suspicious activities in the dependencies that they are monitoring for their customers.}
For example, they may flag dependencies with suspicious fake stars to encourage developers to review their adoption decision and any unusual behaviors or security risks inside the repository. 
To demonstrate the feasibility of suspicious star detection through a third-party tool, we have collaborated with Socket Inc (a) to inform their customers about the threat from fake stars and (b) to integrate our detector to produce alerts for their customers if any of their dependencies have a history of fake star campaigns.
The rationale for this intervention is that the adoption of an open-source component with fake stars deserves stakeholder attention,
and we expect this alert to be used with other supply chain security alerts implemented in the company's product for stakeholder re-evaluation.
At the time of writing, the company's blog post about fake stars had over 2.4k page views, and their tool generated alerts for 283 out of two million distinct customer Software Bill of Materials (SBOMs).
As expected, this alert currently only applies to a small percentage of SBOMs in industry software projects, as we only identified 229 repositories with fake star campaigns and also published in package registries (Section~\ref{sec:rq1}).
Yet, the fact that these packages are found in actual customer SBOMs underscores the practical relevance of our research findings and signals the potential for mitigation through third-party tools.

\emph{Third-party tools can also provide better popularity and trustworthiness signals to replace (or at least supplement) star counts in decision-making.}
For example, there have been a number of ongoing projects that identify security-relevant and activity-based signals for open-source component adoption~\cite[e.g.,][]{DBLP:journals/ieeesp/ZahanKHSW23}. 

\subsection{Researchers}

\emph{With our study, we call for further research on spam and fraudulent activities in the software supply chain.}
While spam and fraud are relatively well studied in emails~\cite{DBLP:journals/air/BlanzieriB08}, social media~\cite{DBLP:journals/snam/AljabriZSASA23}, and e-commerce ~\cite{DBLP:journals/hcisys/HeWCX22}, spam/fraud research on the software supply chain has been scarce (with only a few exceptions~\cite{wu2025exposing, DBLP:conf/acsac/DuYZDWH0020}). 
However, both our findings, and recent gray literature reports~\cite{npm-spam, pypi-spam, docker-hub-spam} indicate that spam and fraudulent activities are rapidly rising in platforms supporting the current software supply chain, including GitHub, npm, PyPI, and DockerHub.
This arguably creates an emerging and important battleground, especially considering their wide usage and critical importance to our modern digital infrastructure~\cite{eghbal2016roads}.
Also, software supply chain attacks are becoming increasingly diverse and sophisticated, evolving from simple typosquatting~\cite{DBLP:conf/dimva/OhmPS020} to complex social engineering attacks as in the \textit{xz} attack~\cite{xz-attack} and reputation farming~\cite{reputation-farm}.
\emph{Although our study does not find any evidence of fake stars being used for social engineering attacks, we call for continuous research and monitoring into this potential threat.}

\section{Conclusion}

In this paper, we have presented a systematic, global, and longitudinal measurement study of fake stars in GitHub.
Our study sheds light on this prevalent and escalating threat happening in a platform central to modern open-source software development. 
Despite its wide prevalence and rising popularity, our findings have also revealed its shady nature: 
Fake stars are, at worst, used to spread malware in short-lived repositories, and at best, used as a short-term promotional tool, which does not bring long-term returns.
Our work emphasizes a critical need for vigilance in assessing other repository signals beyond star counts, and calls for the design of better popularity signals.
Our study also shows a need for further research on spam, fraudulent, and malicious activities in the software supply chain, especially regarding the risk of social engineering attacks.
Future research could expand upon our findings through better fake star detection approaches, exploring additional data (e.g., analyzing fake accounts in GitHub from a social network perspective), and examining similar spam, fraudulent, or malicious activities in other platforms related to software supply chain security.

\section*{Responsible Disclosure}

For all \textsf{Spam/Phishing} repositories we find in RQ3 that were still accessible on GitHub when we discovered them, we reported them immediately to GitHub.
Based on their phishing patterns, we further identified all remaining repositories in our dataset that (1) provide a download link to \Code{.zip}, \Code{.rar} or \Code{.exe} file, (2) VirusTotal~\cite{virus-total} reports the presence of malware, or (3) the README content is highly similar to other repositories marked as \textsf{Spam/Phishing} by our standard.
We manually verified the identified repositories and reported those that align with our open coding criteria for \textsf{Spam/Phishing} repositories in RQ3.
In total, we reported 130 phishing repositories to GitHub, plus two accounts that appear to be almost entirely used for Search Engine Optimization (SEO) spamming (each of them having 213 and 139 repositories, respectively).
At the time of writing (August 2025), all phishing repositories and one of the SEO spamming accounts are no longer accessible in GitHub; the other SEO spamming account only has 17 public repositories remaining.
For the remaining repositories and accounts with suspected fake stars, we do not directly report them, given the possibility that \SystemName may generate false positives and thus false accusations (recall the discussion in Section~\ref{sec:limitations}).
Still, we reached out to the GitHub Security team to inform them of our research---they likely have access to other non-public information, such as IP addresses, that could be used to further validate and refine our detections.

\section*{Data Availability}
\label{sec:data-avail}
\label{sec:conclusion}

We provide the appendix of this paper, the source code of \SystemName, and the data and scripts used in our measurement study at:
\begin{quote}
\centering
\url{https://doi.org/10.5281/zenodo.17009693}
\end{quote}
The appendix is also available in the arXiv version of this paper~\cite{DBLP:journals/corr/abs-2412-13459}.

\section*{Acknowledgement}

He's and Kästner's work was supported in part by the National Science Foundation (award 2206859).
Kapravelos's work was supported in part by the National Science Foundation (award 2207008).
Bogdan's work was supported in part by the National Science Foundation (award 2317168) and in part by research awards from Google and the Digital Infrastructure Fund.
We would also like to thank Socket Inc (\url{https://socket.dev}) and its CEO, Feross Aboukhadijeh, for providing collaboration opportunities, general support, and computational resources for this project.
Additionally, we would like to thank Google Cloud for Researchers program for offering credits to fund the use of Google BigQuery for this research.

\bibliographystyle{ACM-Reference-Format}
\bibliography{references-shorter}

\appendix

\section{GitHub Star Provider Examples}

Table~\ref{tab:fake-star-providers} shows a non-exhaustive list collected in July 2024 through Google Search.
Their prices range from \$0.10 to \$2.00 per star, and they often claim to deliver their stars in days or even hours.
Apart from the examples shown here, a gray literature report indicates that GitHub stars and accounts are also being sold on dark web~\cite{stargazer-goblin}.

\begin{table}[t]
    \small
    \centering
    \caption{A non-exhaustive list of GitHub star merchants}
    \begin{tabular}{llll}
    \toprule
    Provider & Price / Star & Min Amount & (Claimed) Delivery Time \\ 
    \midrule
    Baddhi   & \$0.10 & 100 & $\le$1 week\\
    BuyGitHub  & \$0.12 & 50 & hours \\
    FollowDeh  & \$0.12 & 20 & instant \\ 
    subme.lt & \$0.67 & 100 & hours \\
    Box ID & \$0.21$\sim$\$ 0.41 & 100 & hours \\ 
    R for Rank& \$0.38 & 100 & NA\\ 
    Twidium & \$1.62 & 50 & 7$\sim$30 days\\
    \bottomrule
    \end{tabular}
    \label{tab:fake-star-providers}
\end{table}

\section{Definition of Repository Categories}

\begin{figure}
    \centering
    \includegraphics[width=\linewidth]{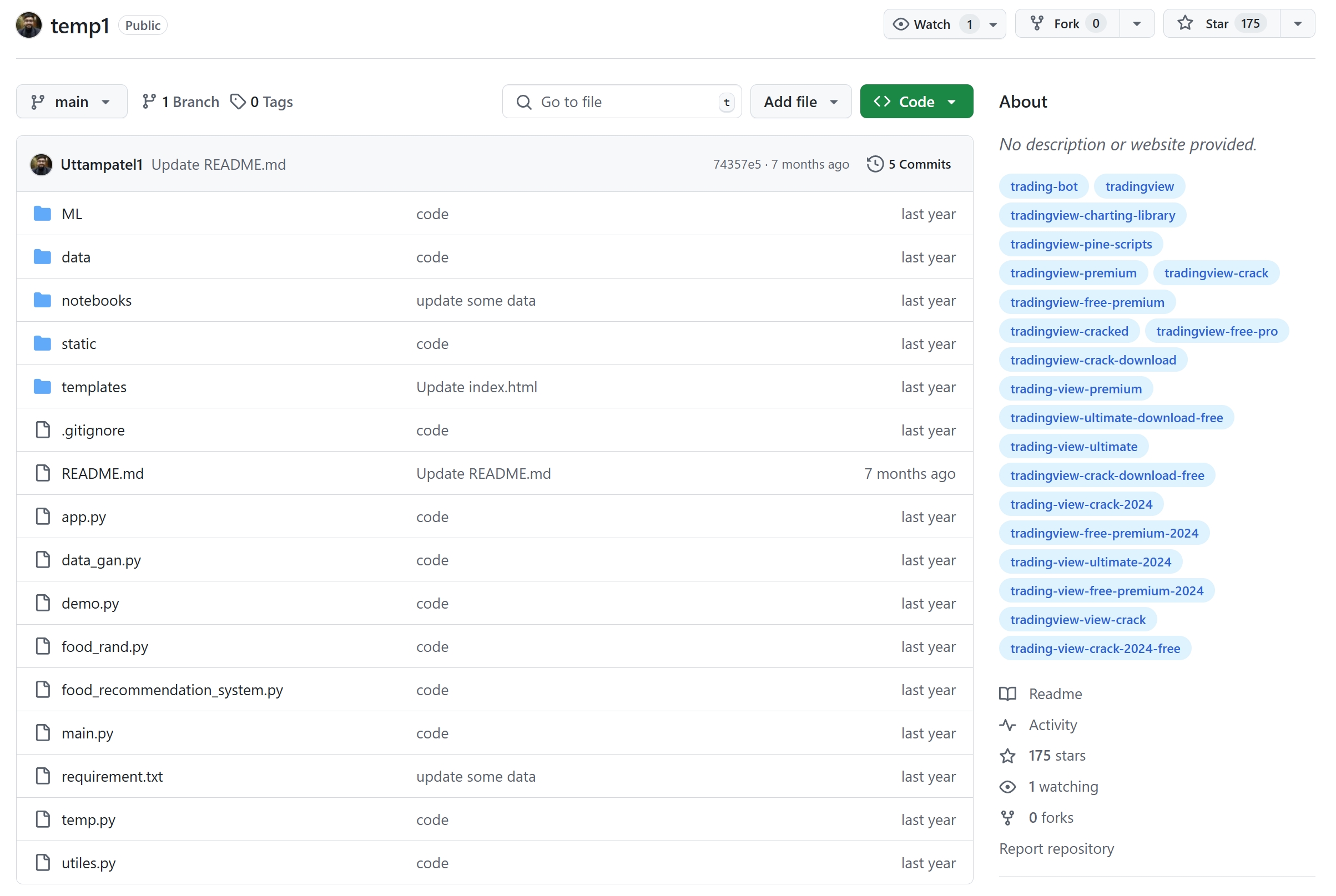}
    \includegraphics[width=\linewidth]{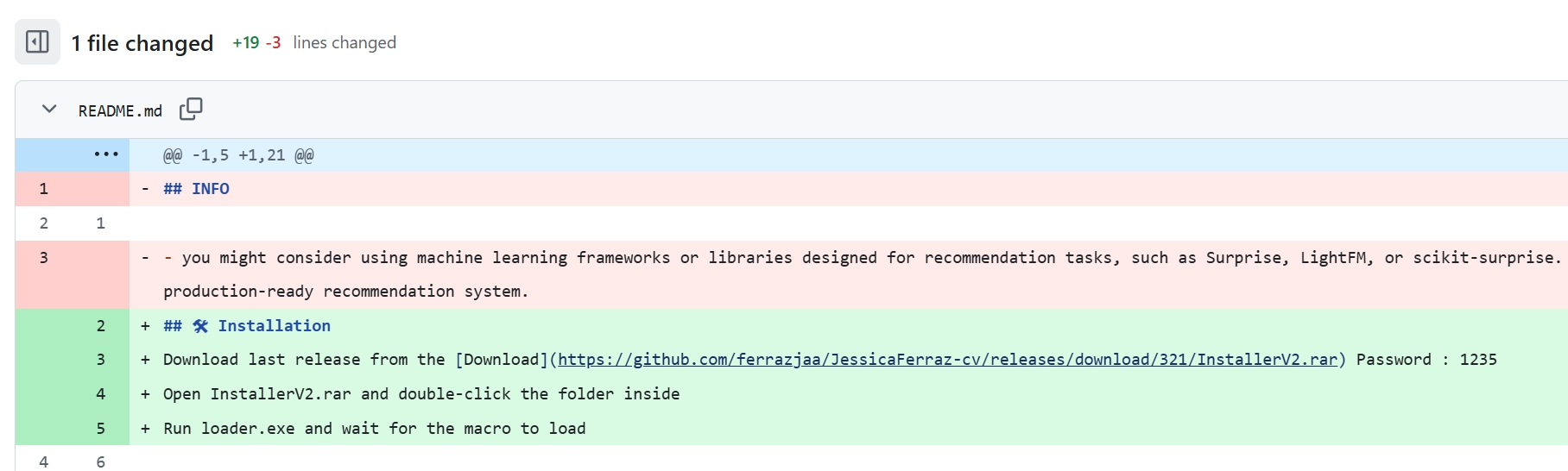}
    \caption{In repository \Code{Uttampatel1/temp1} (197 fake stars as detected by \SystemName, 175 fake stars remaining at the time of writing), its README and repository files initially indicated a machine learning repository. The phishing malware link was only added in a later commit six months after creation. }
    \label{fig:example-malware-later}
\end{figure}

We provide detailed definition of each category as follows:

\begin{itemize}[leftmargin=15pt]
    \item \textsf{Spam/Phishing}: 
    The repository is one of the following cases: (a) it is an obvious clickbait (e.g., claiming game cheats, pirated software, cryptocurrency bots) to trick download of suspicious files, (b) it is spreading spam advertising content, possibly used for search engine optimizations, or (c) it does not contain anything meaningful and thus looks highly suspicious. We conjecture, based on an example we found in Figure~\ref{fig:example-malware-later}, that the third case could be an attacker preparing for a spam/phishing attack.
    \item \textsf{AI/LLM}:
    The repository is generally related to AI or large language models (LLMs). Many of repositories in this category are academic paper repositories or repositories for LLM-related startup products. If the repository appears like a demo, toy, tutorial, or a list of AI-related references (e.g., awesome LLM tools), we put it into the \textsf{Tutorial/Demo} category.
    \item \textsf{Blockchain}: The repository is generally related to blockchain, containing terms related to cryptocurrencies, decentralized finance, or Web 3.0 technologies. Plus, it appears to be a piece of legitimate software. Otherwise, if it appears like a tutorial or list of references, we put it into the \textsf{Tutorial/Demo} category; if it does not appear to contain any meaningful documentation or source code, we put it into the \textsf{Spam/Phishing} category.
    \item \textsf{Web Frameworks}: The repository generally serves the purpose of web development (e.g., platforms for blogging or e-commerce, UI design frameworks, microservice frameworks). Again, if it appears like a demo or template (e.g., a template for \Code{vue.js} development), we put it into the \textsf{Tutorial/Demo} category.
    \item \textsf{Tutorial/Demo}: The repository generally provides reference/educational information for someone else. It may be a list of ``awesome'' other reference repositories, coding tutorials, starter templates for other frameworks, demonstrations, etc.
    \item \textsf{Tool/Application}: The repository is a (seemingly legitimate) piece of tool or application (e.g., command-line tools or Android apps), that are not blockchain-related, AI/LLM-related, or database-related.
    \item \textsf{Basic Utility}: The repository is a software library providing basic utility functions (e.g., caching, exception handling).
    \item \textsf{Database}: The repository is generally related to database. It may be a database solution in itself (e.g., a certain type of database run by a startup), or supporting libraries/tools for databases.
    \item \textsf{Other}: All other repositories that cannot be fit into any of the above categories.
\end{itemize}

\begin{figure}
    \centering
    \includegraphics[width=\linewidth]{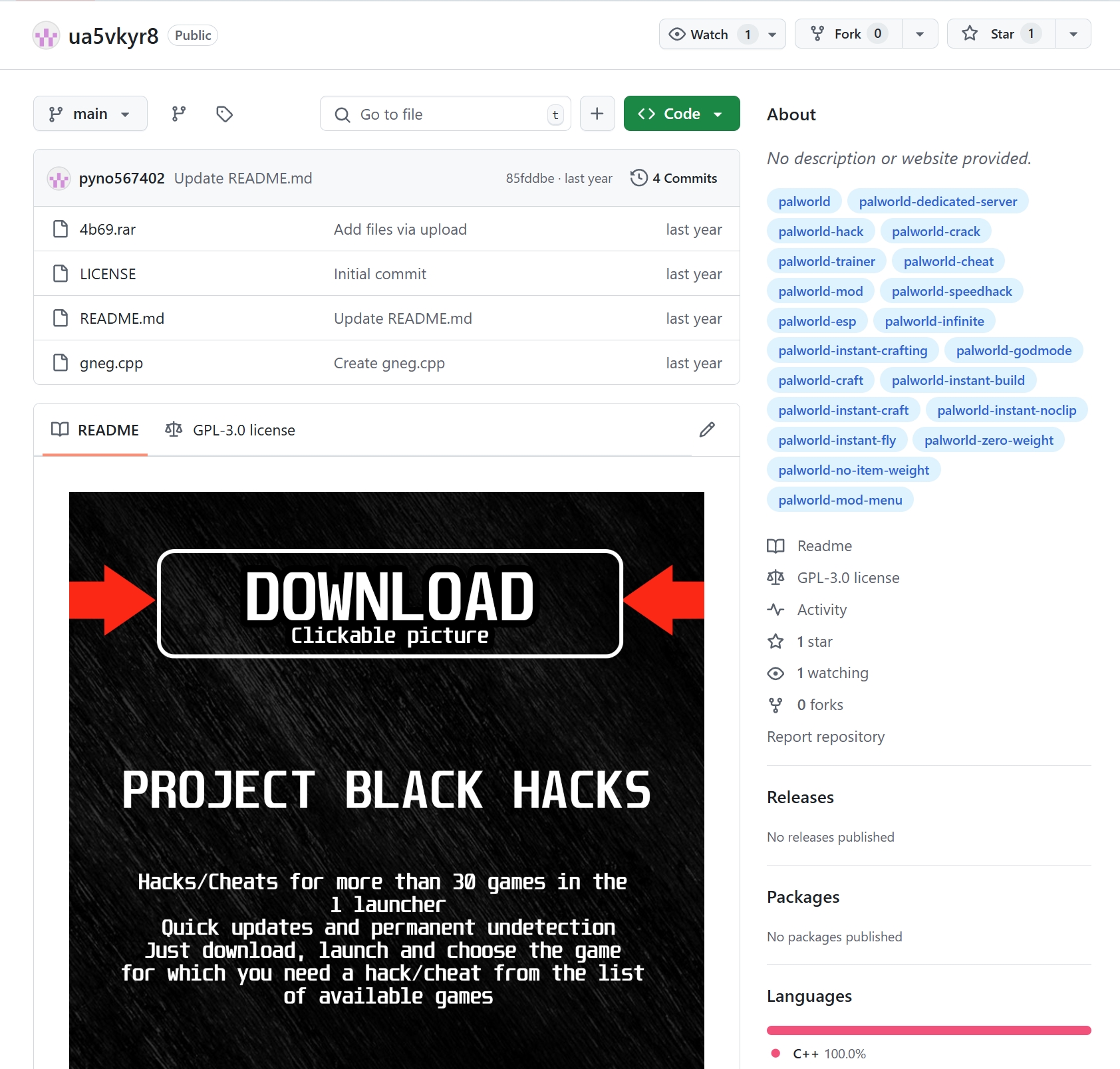}
    \caption{An example phishing repository claiming to be game cheats, with 149 fake stars detected by \SystemName. While the repository is still accessible at the time of writing, almost all of its fake stars have been removed.}
    \label{fig:example-malware-cheats}
\end{figure}

\begin{figure}
    \centering
    \includegraphics[width=\linewidth]{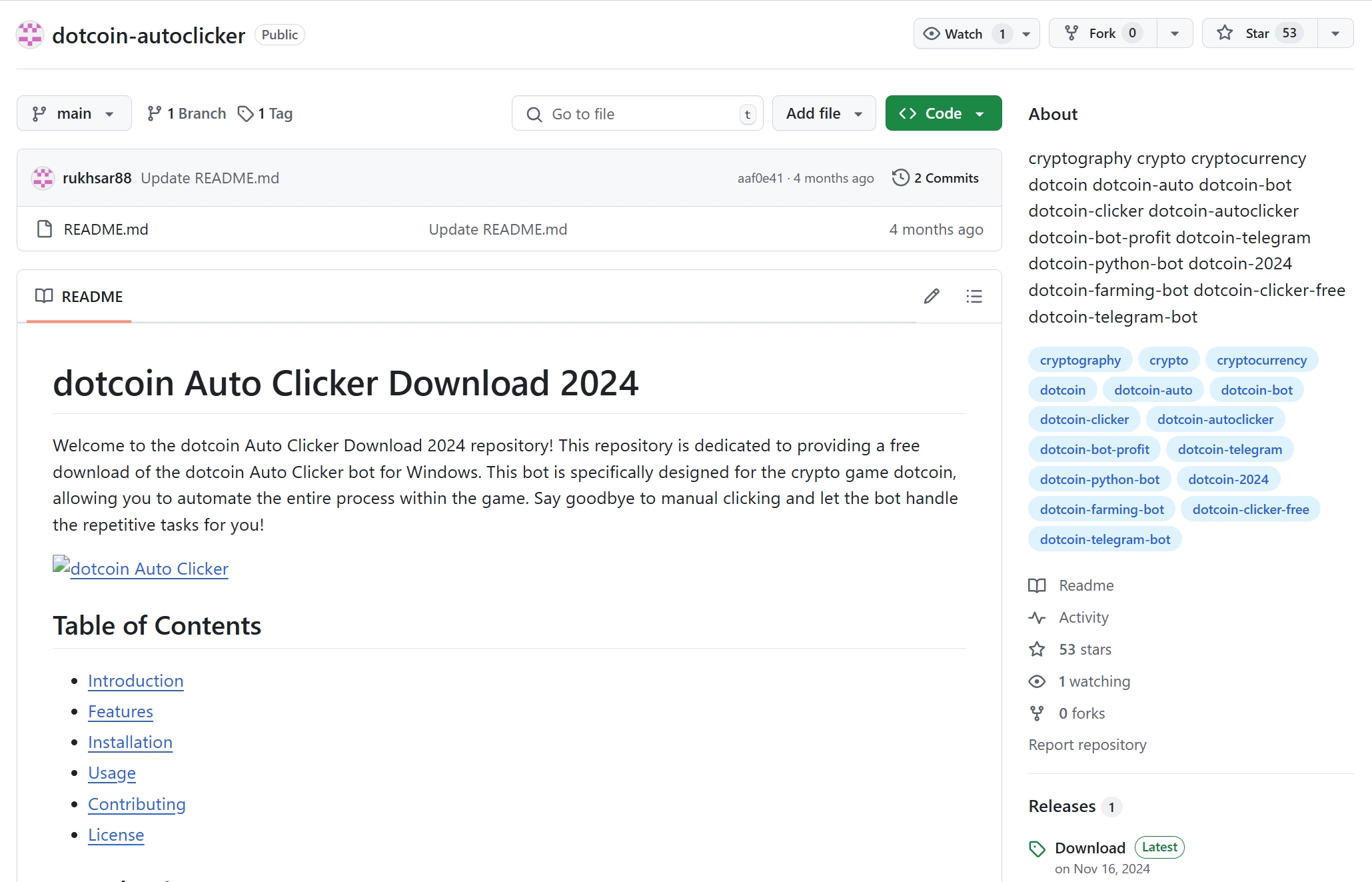}
    \caption{An example phishing repository claiming to be a cryptocurrency bot, with 115 fake stars detected by \SystemName. At the time of writing, the repository still has more than 50 fake stars remaining and visible in its GitHub repository.}
    \label{fig:example-malware-crypto}
\end{figure}

\section{Inter-Rater Reliability Analysis}

In the absence of an available second human coder for the open-coding process we conducted in RQ3, we use GPT-o3-mini, one of the state-of-the-art reasoning LLMs, to act as the second coder.
For each repository, we supply a prompt providing: (a) a description of the open-coding task, (b) the definition of each open-coding category, and (c) the repository README.
The Cohen's $\kappa$ between the human coding results and AI coding results is 0.71, indicating strong agreement~\cite{landis1977measurement}.
Looking over the disagreement cases, we find that GPT-o3-mini, even with built-in reasoning, may fail to recognize better-obfuscated phishing repositories as long as they contain seemingly legitimate READMEs; the remaining disagreements come from confusion between seemingly overlapping labels (e.g., a repository that is an AI/LLM toolkit may be labeled as \textsf{Tool/Application} despite explicit instruction in the prompt not to do so).
Still, LLM's judgements provide a good reference for verification of our open coding results.
We use the LLM's coding results to correct human coding errors before we aggregate and report the open coding results in RQ3.

\section{Spam/Phishing Repository Examples}

We provide two examples of obviously phishing repositories in Figure~\ref{fig:example-malware-cheats} and Figure~\ref{fig:example-malware-crypto}.
Note that the two repositories have different level of fake star retention, indicating that GitHub is actively taking action to counter fake stars.
However, we are also observing evidence indicating that malicious actors are evolving their behaviors to evade platform moderation.
For example, the repository in Figure~\ref{fig:example-malware-later} is initially disguising as a Python machine learning repository; the phishing malware link was only added in a later commit six months after creation.  
It also has a very high level of fake star retention (175 out of 197 fake stars remain even after a year).

\section{Additional Regression Results} 

\begin{table}
    \small
    \centering
    \setlength{\tabcolsep}{5pt}
    \caption{$R^2$ values, Bayesian Information Criterion (BIC), and Akaike Information Criterion (AIC) for regression models with 1-6 autoregression orders and random/fixed effects.}
    \begin{tabular}{lrrrrrr}
    \toprule
         & \multicolumn{3}{c}{Random Effect} & \multicolumn{3}{c}{Fixed Effect} \\
         &  $R^2$ & AIC & BIC & $R^2$ &  AIC & BIC \\
    \midrule
       AR(1) & 0.562 & 34686.50 & 34746.90 & 0.245 & 32947.30 & 32985.00\\
       AR(2) & 0.665 &31274.40 & 31348.90 & 0.268 & 29619.20 & 29671.40\\
       AR(3) & 0.667 & 31170.90 & 31260.30 & 0.270 & 29576.80 & 29643.80\\
       AR(4) & 0.654 & 25794.90 & 25896.90 & 0.285 & 24692.90 & 24773.10\\
       AR(5) & 0.646 & 22744.60 & 23793.70 & 0.297 & 22744.60 & 22838.40\\
       AR(6) & 0.676 & 22284.80 & 22413.60 & 0.293 & 21370.00 & 21477.40\\
    \bottomrule
    \end{tabular}
    \label{tab:ic}
\end{table}

For all the $AR(k)$ panel autoregression models with random or fixed effects, we summarize their $R^2$ values, Bayesian Information Criterion (BIC), and Akaike Information Criterion (AIC) in Table~\ref{tab:ic}.
They are common model selection criteria for regression models, in which $R^2$ measures goodness-of-fit and AIC/BIC penalizes model overfitting.
While in general, models with higher $R^2$ and lower AIC/BIC~\cite{kuha2004aic} are preferred, all twelve models lead to consistent findings regarding our hypotheses, so we only report the $AR(2)$ random- and fixed-effects model in the paper (Table~\ref{tab:regression}) for the sake of simplicity.
The Hausman test~\cite{amini2012fixed} indicate that the unique errors are correlated with the regressors, so results from the fixed-effects model are preferable.
We present the remaining regression results from other $AR(k)$ ($k = 1, 3, 4, 5, 6$) models in appendix Table~\ref{tab:all-regressions}.
We can see that the coefficients of $real_{i, t-k}$ and $all\_real_{i, t-k-1}$ are consistently positive and significant, which supports our \textbf{H$_{1}$}, with only a few exceptions when $k$ is large (Section~\ref{sec:rq4}).
Similarly, $fake_{i, t-1}$ is uniformly significantly positive and $all\_fake_{i, t-k-1}$ is uniformly significantly negative across all ten models, so our key findings around \textbf{H$_{2}$} in Section~\ref{sec:rq4} can be supported using either of the models in our model family.

\begin{table*}
\small
\begin{center}
\caption{Results from $AR(k)$ ($k = 1, 3, 4, 5, 6$) random effect (RE) and fixed effect (FE) panel autoregression models.}
\setlength{\tabcolsep}{5pt}
\begin{tabular}{lcccccccccc}
\toprule 
 & \multicolumn{10}{c}{\textit{Dependent Variable: $real_{it}$}} \\
 \addlinespace
 & \multicolumn{2}{c}{$AR(1)$} &\multicolumn{2}{c}{$AR(3)$} &\multicolumn{2}{c}{$AR(4)$} &\multicolumn{2}{c}{$AR(5)$} & \multicolumn{2}{c}{$AR(6)$}\\
 & RE & FE & RE & FE & RE & FE & RE & FE & RE & FE \\
\midrule
$real_{i, t-1}$      & $0.463^{***}$  & $0.392^{***}$ & $0.479^{***}$  & $0.356^{***}$  & $0.421^{***}$  & $0.343^{***}$  & $0.405^{***}$  & $0.345^{***}$  & $0.421^{***}$  & $0.348^{***}$  \\
                         & $(0.014)$      & $(0.019)$     & $(0.015)$      & $(0.018)$      & $(0.017)$      & $(0.019)$      & $(0.018)$      & $(0.019)$      & $(0.018)$      & $(0.020)$      \\
\addlinespace
$real_{i, t-2}$        &                &               & $0.153^{***}$  & $0.132^{***}$  & $0.181^{***}$  & $0.148^{***}$  & $0.180^{***}$  & $0.153^{***}$  & $0.191^{***}$  & $0.155^{***}$  \\
                         &                &               & $(0.014)$      & $(0.015)$      & $(0.015)$      & $(0.016)$      & $(0.016)$      & $(0.016)$      & $(0.016)$      & $(0.017)$      \\
\addlinespace
$real_{i, t-3}$         &                &               & $0.085^{***}$  & $0.047^{***}$  & $0.108^{***}$  & $0.089^{***}$  & $0.108^{***}$  & $0.098^{***}$  & $0.114^{***}$  & $0.094^{***}$  \\
                         &                &               & $(0.013)$      & $(0.014)$      & $(0.014)$      & $(0.015)$      & $(0.015)$      & $(0.016)$      & $(0.015)$      & $(0.016)$      \\
\addlinespace
$real_{i, t-4}$        &                &               &                &                & $0.059^{***}$  & $0.051^{***}$  & $0.064^{***}$  & $0.052^{***}$  & $0.068^{***}$  & $0.054^{***}$  \\
                         &                &               &                &                & $(0.014)$      & $(0.018)$      & $(0.016)$      & $(0.018)$      & $(0.018)$      & $(0.019)$      \\
\addlinespace
$real_{i, t-5}$          &                &               &                &                &                &                & $0.015$        & $-0.018$       & $-0.006$       & $-0.030^{*}$   \\
                         &                &               &                &                &                &                & $(0.015)$      & $(0.016)$      & $(0.016)$      & $(0.017)$      \\
\addlinespace
$real_{i, t-6}$       &                &               &                &                &                &                &                &                & $0.042^{**}$   & $0.024$        \\
                         &                &               &                &                &                &                &                &                & $(0.019)$      & $(0.019)$      \\
\addlinespace
$all\_real_{i, t-k-1}$  & $0.116^{***}$  & $0.136^{***}$   & $0.047^{***}$  & $0.089^{***}$ &  $0.017^{*}$    & $0.032^{**}$   & $0.016^{*}$    & $0.052^{***}$  &  $-0.005$       & $0.025^{*}$\\
                        & $(0.011)$      & $(0.024)$ &$(0.008)$      & $(0.008)$      & $(0.019)$      & $(0.009)$      & $(0.019)$   & $(0.02)$ & $(0.009)$      & $(0.016)$\\
\addlinespace
$fake_{i, t-1}$         & $0.101^{***}$  & $0.093^{***}$ & $0.063^{***}$  & $0.077^{***}$  & $0.081^{***}$  & $0.084^{***}$  & $0.092^{***}$  & $0.079^{***}$  & $0.087^{***}$  & $0.077^{***}$  \\
                         & $(0.010)$      & $(0.011)$     & $(0.011)$      & $(0.013)$      & $(0.012)$      & $(0.013)$      & $(0.012)$      & $(0.013)$      & $(0.013)$      & $(0.013)$      \\
\addlinespace
$fake_{i, t-2}$        &                &               & $0.011$        & $0.019^{*}$    & $0.002$        & $0.008$        & $0.014$        & $0.015$        & $0.015$        & $0.017$        \\
                         &                &               & $(0.010)$      & $(0.011)$      & $(0.013)$      & $(0.014)$      & $(0.013)$      & $(0.014)$      & $(0.014)$      & $(0.015)$      \\
\addlinespace
$fake_{i, t-3}$        &                &               & $0.027^{**}$   & $0.034^{**}$   & $-0.007$       & $-0.013$       & $-0.014$       & $-0.018$       & $-0.018$       & $-0.014$       \\
                         &                &               & $(0.011)$      & $(0.014)$      & $(0.014)$      & $(0.016)$      & $(0.014)$      & $(0.016)$      & $(0.015)$      & $(0.017)$      \\
\addlinespace
$fake_{i, t-4}$         &                &               &                &                & $-0.014$       & $-0.001$       & $-0.010$       & $-0.005$       & $-0.009$       & $-0.002$       \\
                         &                &               &                &                & $(0.013)$      & $(0.019)$      & $(0.017)$      & $(0.019)$      & $(0.018)$      & $(0.020)$      \\
\addlinespace
$fake_{i, t-5}$         &                &               &                &                &                &                & $-0.036^{**}$  & $0.008$        & $0.008$        & $0.024$        \\
                         &                &               &                &                &                &                & $(0.016)$      & $(0.019)$      & $(0.021)$      & $(0.022)$      \\
\addlinespace
$fake_{i, t-6}$        &                &               &                &                &                &                &                &                & $-0.099^{***}$ & $-0.095^{***}$ \\
                         &                &               &                &                &                &                &                &                & $(0.023)$      & $(0.025)$      \\
\addlinespace
$all\_fake_{i, t-k-1}$  & $0.011^{*}$    & $-0.020^{**}$  & $-0.037^{***}$ & $-0.065^{***}$ & $-0.034^{***}$ & $-0.062^{***}$ & $-0.039^{***}$ & $-0.109^{***}$  & $-0.015^{*}$   & $-0.073^{***}$ \\
                         & $(0.007)$      & $(0.014)$ & $(0.006)$      & $(0.015)$      & $(0.006)$      & $(0.015)$  & $(0.007)$      & $(0.017)$  & $(0.006)$      & $(0.014)$        \\
\addlinespace
$age_{i,t}$                      & $-0.005^{***}$ &               & $-0.003^{***}$ &                & $-0.001$       &                & $-0.000$       &                & $-0.001$       &                \\
                         & $(0.001)$      &               & $(0.001)$      &                & $(0.001)$      &                & $(0.001)$      &                & $(0.001)$      &                \\
\addlinespace
$release_{i,t}$         & $0.182^{***}$  &               & $0.090^{***}$  &                & $0.067^{***}$  &                & $0.073^{***}$  &                & $0.047^{**}$   &                \\
                         & $(0.030)$      &               & $(0.021)$      &                & $(0.023)$      &                & $(0.024)$      &                & $(0.022)$      &                \\
\addlinespace
$activity_{i,t}$        & $0.166^{***}$  &               & $0.086^{***}$  &                & $0.085^{***}$  &                & $0.084^{***}$  &                & $0.071^{***}$  &                \\
                         & $(0.012)$      &               & $(0.010)$      &                & $(0.010)$      &                & $(0.010)$      &                & $(0.010)$      &                \\
\midrule
Observations               & $14,019$       & $14,019$      & $12,738$       & $12,738$       & $10,798$       & $10,798$       & $10,054$       & $10,054$       & $9,469$        & $9,469$        \\
R$^2$                    & $0.562$        & $0.245$       & $0.667$        & $0.270$        & $0.654$        & $0.285$        & $0.646$        & $0.297$        & $0.675$        & $0.293$        \\
Adjusted R$^2$              & $0.562$        & $0.169$       & $0.667$        & $0.205$        & $0.654$        & $0.231$        & $0.646$        & $0.252$        & $0.674$        & $0.252$        \\

\bottomrule
\textit{Note:}  & \multicolumn{10}{r}{$^{*}$p$<$0.1; $^{**}$p$<$0.05; $^{***}$p$<$0.01} \\ 
\end{tabular}

\label{tab:all-regressions}
\end{center}
\end{table*}

\end{document}